\documentclass{cmslatex}

\usepackage[paperwidth=7in, paperheight=10in, margin=.875in]{geometry}
\usepackage{graphics}
\usepackage{hyperref}
\usepackage{amsmath}
\usepackage{amssymb} 
\usepackage{amsfonts}
\usepackage{mathtools}
\usepackage{enumitem}
\usepackage{color,url}
\usepackage{bbm} 
\usepackage[normalem]{ulem}

\definecolor{ggreen}{cmyk}{1,     0,      1,      0}
\definecolor{orang}{rgb}{1, 0.40001, 0.2}
\definecolor{viol}{rgb}{0.6, 0.2, 0.9}
 
\renewcommand{\theequation}{\arabic{section}.\arabic{equation}}

\newtheorem{thm}{Theorem}[section]
\newtheorem{prop}[thm]{Proposition}
\newtheorem{lem}[thm]{Lemma}
\newtheorem{cor}[thm]{Corollary}
\newtheorem{propt}[thm]{Property}

\newtheorem{defn}[thm]{Definition}

\newtheorem{rem}[thm]{Remark}

\newcommand{\dps}{\displaystyle}

\newcommand{\RR}{\mathbb{R}}

\newcommand{\Splus}{\ensuremath{\mathbb{S}^2_+}}

\begin{document}
\title{From the simple reacting sphere kinetic model to 
        the reaction-diffusion system of Maxwell-Stefan type
        }

\author{
          Benjamin Anwasia\thanks{Centro de Matem\'atica, Universidade do Minho, Braga, Portugal,         
          \url{id6226@alunos.uminho.pt}}
          \and {Patr\'icia Gon\c{c}alves}\thanks{Center for Mathematical Analysis, Geometry and Dynamical Systems, 
          Instituto Superior T\'ecnico, Universidade de Lisboa, Portugal,
          \url{patricia.goncalves@math.tecnico.ulisboa.pt}}
          \and{Ana Jacinta Soares}\thanks{Centro de Matem\'atica, Universidade do Minho, Braga, Portugal,  
          \url{ajsoares@math.uminho.pt}}
          }


\pagestyle{myheadings} 
\markboth{From the SRS kinetic model to reaction diffusion equations of MS type}
                {B. Anwasia, P. Gon\c{c}alves, A.J. Soares} 
\maketitle
          

\begin{abstract}
In this paper we perform a formal asymptotic analysis on a kinetic model for reactive mixtures in order to derive 
a reaction-diffusion system of Maxwell-Stefan type. 
More specifically, we start from the kinetic model of simple reacting spheres for a quaternary mixture of 
monatomic ideal gases that undergoes a reversible chemical reaction of bimolecular type. 
Then, we consider a scaling describing a physical situation in which 
mechanical collisions play a dominant role in the evolution process, 
while chemical reactions are slow, 
and compute explicitly the production terms associated to 
the concentration and momentum balance equations for each species in the reactive mixture. 
Finally, we prove that, under     
isothermal assumptions,
the limit equations for the scaled kinetic model is the reaction diffusion system of Maxwell-Stefan type. 
\end{abstract}
          




                
\section{Introduction}
\label{intro}

The description and modelling of chemically reactive mixtures is a topic of great importance due to many engineering applications related, for example, to chemical industry and biotechnology \cite{giovangigli, TK93,WK06}. 
In particular, a proper description of diffusive phenomena in multicomponent mixtures, with or without chemical reactions, is crucial in many simulations and design processes used by chemical engineers \cite{KW97}.
In this context, the Maxwell-Stefan (MS) equations are used by many applied and experimental researchers 
to model and predict diffusion as well as mass transfer processes in multicomponent mixtures \cite{TK93, BK95,WK06}.
In fact, it is well known that the MS equations are adequate to describe non-typical diffusions 
that appear as a consequence of some thermodynamic non-idealities,
by introducing the chemical potential gradients as driving forces
\cite{KW97,WK06}. 
When a multicomponent mixture with chemical reaction is considered, 
a hydrodynamic system which consists of the continuity equations for the constituents in the mixture and the MS equations, 
can be used to describe diffusion among the constituents and how their concentrations 
change as a consequence of the chemical reaction.

Despite the practical interest and applications of the MS equations
for multicomponent mixtures with or without chemical reaction, 
not much is known about the mathematical analysis of these equations. 
Rigorous results have been published over the past few years, 
see \cite{Bothe11, BoudinGS12, BoudinGS15, BGP17, Herberg16, HS17, Jungel13}.
In particular, \cite{BoudinGS15, BGP17, HS17} dealt with the formal derivation of 
hydrodynamic systems of MS equations coupled with the continuity equations for the species
from a kinetic (mesoscopic) system of Boltzmann equations for a non-reactive 
multicomponent mixture and obtained explicit expressions for the diffusion coefficients in terms of the kinetic model parameters.
Such papers follow the well established line of research initiated by Golse and co-workers \cite{BardosGL89, BardosGL91,BardosGL93} 
on the transition from kinetic Boltzmann models to hydrodynamic equations of fluid mechanics.
In close connection with these works, but considering chemically reactive mixtures, 
we quote here  \cite{BisiD06, BisiD08, BisiS06} which dealt with the derivation of 
macroscopic reaction-diffusion equations from a system of reactive Boltzmann equations. 
In these papers, using an appropriate scaling of the reactive Boltzmann equations 
and assuming different types of molecular interactions, 
the evolution equations for the species number densities were explicitly derived in the asymptotic limit of small Knudsen number. 
Moreover, the convergence from the reactive Boltzmann equations to 
the reaction-diffusion system was proven and discussed in \cite{BisiD06, BisiD08}.
Reaction-diffusion equations for chemically reactive mixtures were also derived in  \cite{SZ92, SZ93,Z93},
starting from kinetic equations of Fokker-Planck type. 
Various scalings were considered in view of analyzing the interactions between transport processes and chemical reactions. 

In the present paper, we are interested in the limiting process that leads from a particular model of 
reactive Boltzmann equations to a reaction-diffusion system of MS type. 
More precisely, we consider a quaternary mixture of monatomic ideal
gases undergoing a bimolecular reversible reaction, 
described by the simple reacting sphere (SRS) kinetic model 
\cite{PS17, QinD95, XystrisD78},
in which both elastic and reactive collisions are of hard-sphere type.
Then, considering a scaling of the SRS kinetic equations for which elastic collisions are dominant 
and reactive collisions are less frequent, 
and assuming isothermal conditions, 
we formally derive the reaction-diffusion system of MS type for the evolution of the number density and momentum of each species.
The formal derivation of these equations from the SRS kinetic model is our main contribution in this paper and, as far as we know, this is the first attempt in this direction.

After this introduction, the remaining part of our work is organized as follows. 
The reaction-diffusion system of MS type  for a multi-species reactive mixture in the 
context of continuum mechanics is introduced in Section \ref{sec:ms}. 
In Section \ref{sec:srs} we describe the SRS kinetic model and
introduce its relevant properties that are essential for our analysis. 

The diffusive asymptotics of the SRS kinetic model towards the reaction-diffusion system of MS type is studied in Section \ref{sec:rdl}, after a proper scaling of the SRS equations. Our conclusions and some future perspectives are stated in Section \ref{sec:conc}. Finally, we include an Appendix in Section \ref{sec:App} where we give some steps and other details about the computation of the integrals appearing in Section \ref{sec:rdl}.

\medskip


\section{The continuum reaction-diffusion system of Maxwell-Stefan type}
\label{sec:ms}

In this section, we introduce a mathematical model for a reactive multi-species gaseous mixture in the context of continuum mechanics.
The mixture is influenced by two processes, 
namely the diffusion, which causes the species to spread in space, 
and a chemical reaction, which results in the transformation of the species into each other.
The model equations consist of the concentration balance equations for the reactive species in the mixture 
coupled with the MS equations for the momentum of the species.
These equations describe how both processes affect the evolution of the mixture
and will be referred to as the {\it reaction-diffusion system of MS type}. 

Let $\Omega \subset {\mathbb{R}}^3$ be a bounded domain with boundary denoted by $\partial\Omega$ and outward normal vector at each point ${\bf{x}}$ of the boundary given by $\nu({\bf{x}})$. 
We consider a mixture of four species, say $A_1, A_2, A_3$ and $A_4$,  that participate in a chemical reaction of  type
\begin{equation}
A_1+A_2\rightleftharpoons A_3+A_4.
\label{eq:reac} 
\end{equation}
This means that species $A_1,A_2$ react to produce species $A_3,A_4$ and conversely, species $A_3,A_4$  also react to produce species $A_1,A_2$.
We say that $A_1,A_2$ and $A_3,A_4$ are the reactive species (or the reactive pairs),
more specifically $A_1,A_2$ are reactants and $A_3,A_4$ are products of the forward chemical reaction. 
For each species $A_i$, with $i=1,2,3,4$, 
let $\varrho_i(t,{\bf{x}})\geq 0$ be the mass density,   
${\bf{u}}_i(t,{\bf{x}})$ the mean velocity and  
$r_i(t,{\bf{x}})$ the production rate of mass density due to the chemical reaction,
where ${\bf{x}}\in\Omega$ and $t>0$. 

The mass balance equation for each constituent in the reactive mixture reads
\begin{equation}
\frac{\partial \varrho _i}{\partial t}+\frac{\partial }{\partial {\bf{x}}}(\varrho _i{\bf{u}}_i)=r_i,\quad   {\bf{x}}\in\Omega,\quad t>0.
\label{eq:massbalance}
\end{equation}
Due to the type of chemical reaction \eqref{eq:reac}, 
the production rates satisfy the condition
\begin{equation}
\sum_{i=1}^{4} r_i=0,
\label{eq:r}
\end{equation}
which results in the conservation of the total mass of the mixture.

In this paper, we adopt a molar based description (see \cite{Do98}) of the reactive mixture and, 
as a result, for each species $A_i$ we introduce 
the molar mass $M_i$
and define
the molar concentration $c_i(t,{\bf{x}})$ and its production rate ${\cal J}_i$, 
the molar flux ${\bf{N}}_i(t,{\bf{x}})$, 
the mole fraction $\gamma_i(t,{\bf{x}})$ and 
the molar diffusive flux ${\bf {J}}_i(t,{\bf x})$,
given respectively by
\begin{equation}
c_i = \frac{\varrho_i}{M_i}, \qquad
{\cal J}_i = \frac{r_i}{M_i}, \qquad
{\bf{N}}_i = c_i {\bf{u}}_i, \qquad 
\gamma_i =\frac{c_i }{c }, \qquad
{\bf{J}}_i =c_i({\bf{u}}_i - {\bf{v}}) .
\label{eq:ciNigammaiji}
\end{equation}
Here, $c \coloneqq c(t,{\bf{x}})$ and ${\bf v} \coloneqq {\bf v} (t,{\bf x})$ are the molar concentration and 
molar average velocity of the mixture,
defined by
\begin{equation}
c = \sum_{i=1}^{4}c_i, \qquad 
{\bf{v}}= \sum_{i=1}^{4} \gamma_i{\bf{u}}_i,  \qquad {\bf{x}}\,\in\,\Omega,\quad t>0.
\label{eq:totalcon_velocity }
\end{equation}
Note that the molar diffusive fluxes ${\bf{J}}_i$ satisfy the constraint
\begin{equation}
\sum_{i=1}^{4}{\bf{J}}_i=0,  \qquad {\bf{x}}\,\in\,\Omega,\quad t>0.
\label{eq:Jsum}
\end{equation}

In our analysis, diffusion and chemical reaction are the relevant effects in the mixture. 
Therefore we neglect the effects due to temperature gradients,
by assuming isothermal conditions which correspond to a uniform in space and constant in time 
mixture temperature $T$.
We also neglect the effects due to convection and advection.
%
%
%

From the mass balance equation \eqref{eq:massbalance}, using definitions \eqref{eq:ciNigammaiji},
we obtain the evolution equations for the species concentrations $c_i$ in the form
\begin{equation}
\frac{\partial c_i}{\partial t}+\frac{\partial {\bf {J}}_i}{\partial {\bf{x}}} = {\cal J}_i ,
\quad {\bf{x}}\,\in\,\Omega, \quad t>0,\quad i=1,2,3,4,
\label{eq:RDE}
\end{equation}
where we have considered $\frac{\partial}{\partial \bf x}(c_i \bf v)=0$,
since convection is neglected.

\medskip

The diffusion process in multicomponent gaseous mixtures can be accurately 
described by the MS equations,
which express the relationship between the molar diffusive fluxes and concentrations of the 
chemical potentials of the species.
Adopting the standard form for the chemical potentials \cite{Herberg16}
and taking into account the isothermal assumptions, 
the driving forces become proportional to the concentration gradients of the species 
and the MS equations relate the molar diffusive fluxes to the concentration gradients. 
Following the description of \cite{TK93}, and using our notation,
the MS equations under the isobaric assumption (i.e. with constant pressure) can be written in the form
\begin{equation}
\frac{\partial c_i}{\partial \bf x} = -\frac{1}{c} \sum_{\substack{s=1\\ s\neq i}}^{4}
        \frac{c_s{\bf J}_i - c_i {\bf J}_s}{ D_{is}}, \quad {\bf{x}}\,\in\,\Omega,\quad t>0,\quad i=1,2,3,4,
\label{eq:MSE}
\end{equation}
where $D_{is}$ is the diffusion coefficient associated to species $A_i$ and $A_s$, 
with $D_{is}=D_{si}$.

\medskip

Observe that summing \eqref{eq:RDE} over all species  
as well as \eqref{eq:MSE} over all species, we obtain
$$
\frac{\partial c}{\partial t}  = 0 
\qquad \mbox{and} \qquad
\frac{\partial c}{\partial \bf x}  = 0,
$$
which means that the total molar concentration of the mixture, $c$, 
is uniform in $x$ and constant in time.

Observe also that the MS equations \eqref{eq:MSE} are linearly dependent and only three of these equations are independent, so we have to add another equation to the system.  

\medskip

Equations \eqref{eq:RDE} and \eqref{eq:MSE}, together with constraint \eqref{eq:Jsum},
constitute the reaction-diffusion system of MS type.
It describes the diffusion and chemical kinetics of the multi-species reactive mixture in the context of continuum mechanics.

For what concerns the boundary conditions to join to our set of equations, 
we assume that the chemical reaction \eqref{eq:reac} takes place in a {closed domain},
so that we impose
\begin{equation}
\nu \cdot {\bf{J}}_i = 0, \qquad 
{\bf{x}}\in \partial\Omega, \qquad t>0, \qquad i=1,2,3,4.
\label{eq:BDcondition}
\end{equation}

The aim of the present paper is to formally derive the balance equations \eqref{eq:RDE} and the MS equations \eqref{eq:MSE}
as the hydrodynamic limit of the SRS kinetic model for the considered reactive mixture.
The chemical production rates  ${\cal J}_i$ and the diffusion coefficients $D_{is}$
will be explicitly computed from the collisional dynamics of the kinetic model
and will be expressed in terms of some kinetic parameters.

\bigskip


\section{The SRS kinetic model}
\label{sec:srs}

In this section, we introduce our kinetic model for the quaternary reactive mixture 
considered in Section \ref{sec:ms}.
This model is based on the kinetic theory of simple reacting spheres (SRS),
first proposed by Marron in \cite{Marron70}, and then developed by Xystris, Dahler and Qin in 
\cite{DQ03, QinD95, XystrisD78}. 
Some aspects of the mathematical analysis of the SRS model were investigated, for example, in 
\cite{CarvalhoPhD13, GP04, Polewczak00,PS17}. 
Here, we introduce the model and briefly describe some of its properties needed for the analysis
developed in this paper. 
Other details about the SRS model can be seen in the references just quoted above.

\medskip

We consider the quaternary reactive mixture introduced in Section \ref{sec:ms},
whose constituents $A_1,A_2,A_3,A_4$ 
participate in the bimolecular chemical reaction \eqref{eq:reac}.
Internal degrees of freedom associated to rotational, vibrational and nuclei energies are not taken into account.
For each $i=1,2,3,4$, let $m_i$, $d_i$ and $E_i$ be, respectively, the mass, the diameter and the formation energy of the species $A_i$. 
Conservation of mass holds for the chemical reaction  \eqref{eq:reac} and thus we have
\begin{equation}
m_1+m_2 = m_3+m_4 = M.
\label{eq:masscon}
\end{equation}
The reaction heat, denoted by $Q_R$, is given by the difference between the formation energies of the products of the forward reaction and those of the reactants, i.e.
\begin{equation}
Q_R=E_3+E_4-E_1-E_2 .
\label{eq:Reactionheat}
\end{equation}
This means that the forward reaction $A_1+A_2 \rightarrow A_3+A_4$ is exothermic if $Q_R < 0$.
Other\-wise, it is endothermic.
Also, we introduce the activation energy $\zeta_i$ for each of the species $A_i$, 
such that $\zeta_1=\zeta_2$, $\zeta_3=\zeta_4$ and 
$\zeta_3=\zeta_1 - Q_R$.

\medskip


\subsection{Collisional dynamics} 
\label{ssec:elc}

Particles in the mixture undergo binary elastic collisions and reactive encounters obeying the chemical law \eqref{eq:reac}, 
both of hard sphere type. 
Elastic collisions take place between particles of the same species,
as well as between particles of different species.
The cross section of an elastic collision between particles of species $A_i$, $A_s$  is defined by
\begin{equation}
\sigma _{is}^2=\frac{1}{4}(d_i+d_s)^2,  \qquad i,s=1,2,3,4.
\label{eq:ECS}
\end{equation}
If ${\bf{v}}_i$, ${\bf{v}}_s$ are the pre-collisional velocities   
and ${\bf{v}}_i'$, ${\bf{v}}_s'$ the post-collisional velocities, 
the conservation laws of linear momentum and kinetic energy are,
respectively, given by
             \begin{equation}
             m_i{\bf{v}}_i+m_s{\bf{v}}_s = m_i{\bf{v}}_i'+m_s{\bf{v}}_s' 
             \quad   \mbox{and} \quad
             m_i({\bf{{v}}}_i)^2+m_s({\bf{v}}_s)^2 = m_i({\bf{v}}_i')^2+m_s ({\bf{v}}_s')^2 .
             \label{eq:ECOMAK}
             \end{equation}
The post-collisional velocities are given in terms of the pre-collisional velocities by
             \begin{equation}
             {\bf{v}}_i' = {\bf{v}}_i-2\frac{\mu_{is}}{m_i} \epsilon \left \langle \epsilon ,
             {\bf{v}}_i - {\bf{v}}_s \right \rangle 
             \quad   \mbox{and} \quad 
             {\bf{v}}_s'= {\bf{v}}_s+2\frac{\mu_{is}}{m_s}\epsilon 
             \left \langle \epsilon ,{\bf{v}}_i-{\bf{v}}_s \right \rangle,
             \label{eq:EPCV}
             \end{equation}
where $ \mu_{is}$ is the reduced mass of the colliding pair and is given by
             \begin{equation}
              \mu_{is}=\frac{m_im_s}{m_i+m_s},
              \label{eq:RM}
             \end{equation}
and $\epsilon$ is a unit vector directed along the line joining the centre of the two spheres at the moment of impact, 
that is $\epsilon \in \Splus \!=\! \big\{ \overline\epsilon\in \mathbb{R}^3: \; \| \overline\epsilon\| = 1, \; \langle\overline\epsilon,{\bf{v}}_i - {\bf{v}}_s\rangle > 0\big\}$.
Moreover, $\left\langle \cdot , \cdot \right\rangle$ represents the inner product in $\RR^3$ and
$\|\cdot\|$ is  the norm induced by this inner product.
For convenience, we introduce the total mass of the colliding pair,
$M_{is}=m_i+m_s$.

\medskip

\begin{rem}
\label{rm:sphe}
Note that if we use spherical coordinates with 
$\theta \in [0 , \pi/2 ]$ as the polar angle between ${\bf{v}}_i-{\bf{v}}_s$ and $\epsilon$
and $\phi \in [0 , 2\pi [$ as the azimuthal angle in the plane orthogonal to ${\bf{v}}_i-{\bf{v}}_s$,
then
$\dps \epsilon = \left ( \sin\theta \cos\phi,  \, \sin\theta \sin\phi, \, \cos\theta  \right )$,
$\dps \left \langle \epsilon,{\bf{v}}_i-{\bf{v}}_s \right \rangle  = V \cos\theta$,
where $V= \|{\bf{v}}_i-{\bf{v}}_s\|$ is the relative velocity before the collision. 
\end{rem}

\medskip

Concerning reactive encounters, a collision between particles of species $A_i$, $A_j$
with pre-collisional velocities ${\bf{v}}_i$, ${\bf{v}}_j$  will result in a chemical reaction if the kinetic energy associated with the relative motion of the colliding pair along the line of their centres is greater than, or equal to, the activation energy, that is
\begin{equation}
\frac{1}{2} \mu_{ij}\left \langle \epsilon , {\bf{v}}_i-{\bf{v}}_j \right \rangle^2\geq {\zeta _{i}}.
\label{eq:nccr}
\end{equation}
If $A_k$, $A_l$ represent the products of the forward reaction and ${\bf{v}}_k^{\circ}$, ${\bf{v}}_l^{\circ}$ their
post-collisional velocities, then the conservation laws of linear momentum and total energy (kinetic plus binding)
for reactive collisions are, respectively, given by
              \begin{align}
              m_i{\bf{v}}_i+m_j{\bf{v}}_j  &=  m_k{{\bf{v}}_k}^{\circ}+m_l{{\bf{v}}_l}^{\circ},
              \label{eq:RCM}\\
              E_i+\frac{1}{2}m_i({{\bf{v}}}_i)^2+E_j+\frac{1}{2}m_j({\bf{v}}_j)^2  &= 
              E_k+\frac{1}{2}m_k({\bf{{v}}}_k^{{\circ}})^2+E_l+\frac{1}{2}m_l({\bf{{v}}}_l^{{\circ}})^2 ,
              \label{eq:RCTE}
              \end{align}
where the indexes are such that
$ (i,j,k,l) \!\in\! \left \{ \! (1,2,3,4),(2,1,4,3),(3,4,1,2),(4,3,2,1) \! \right \}\!$. \!
From now on, if nothing is said about the indexes $(i,j,k,l)$, we assume that they 
are as introduced above.

\medskip

\begin{rem}
\label{rem:reaction_condition}
From condition \eqref{eq:nccr}, for a reactive collision to occur, we must have
$\dps \left \langle \epsilon , {\bf{v}}_i-{\bf{v}}_j \right \rangle\geq \sqrt{{2\zeta _{i}}/{\mu_{ij}}} $.
Using the definitions of the unit vector $\epsilon$ 
and relative velocity $V= \|{\bf{v}}_i-{\bf{v}}_j\|$ before the collision, 
we obtain
$\dps V \geq \sqrt{{2\zeta _{i}}/{\mu_{ij}}} $,
which motivates the definition of the threshold relative velocity as
$\dps \Xi_{ij} = \sqrt{{2\zeta _{i}}/{\mu_{ij}}}  
$.
In particular, $\Xi_{ij}$ is the required relative velocity necessary
to assure that the collision will be of reactive type.  
\end{rem}

\medskip

The reactive cross sections for the direct and reverse chemical reactions 
can be defined in terms of their threshold relative velocities
by
\begin{equation}
{\sigma }'^2 _{12}=
\left\{\begin{array}{ll}
\beta _{12}\sigma^2 _{12}, &
\left \langle \epsilon ,{\bf{v}}_1-{\bf{v}}_2 \right \rangle\geq \Xi _{12}, \\[1em]
              0, &\left \langle \epsilon ,{\bf{v}}_1-{\bf{v}}_2 \right \rangle < \Xi _{12},
              \end{array} \right.
              \qquad                
{\sigma }'^2 _{34}=
\left\{\begin{array}{ll}
\beta_{34}\sigma^2 _{34}, &  
\left \langle \epsilon ,{\bf{v}}_3-{\bf{v}}_4 \right \rangle\geq \Xi _{34},\\[1em]
              0, & \left \langle \epsilon ,{\bf{v}}_3-{\bf{v}}_4 \right \rangle < \Xi _{34},
             \end{array}\right.
             \label{eq:RCS}
\end{equation}
where the coefficients $\beta_{ij}$
represent the fraction of colliding pairs with enough 
kinetic energy to produce a reaction that in fact react chemically.
They play the role of steric factors,
with $0\leq\beta_{ij}\leq 1$.

\medskip

\noindent
The post-collisional velocities for the forward chemical reaction 
$A_1+A_2 \rightarrow A_3+A_4$ 
are given by
\begin{equation}
\begin{aligned}
{\bf{v}}_3^{\circ}&=\frac{1}{M}\left [ m_1{\bf{v}}_1+m_2{\bf{v}}_2+ 
                            m_4\sqrt\frac{\mu_{12}}{\mu_{34}}\left \{ ({\bf{v}}_1-{\bf{v}}_2)-
                            \epsilon\left \langle \epsilon,({\bf{v}}_1-{\bf{v}}_2) \right \rangle +
                            \epsilon \omega ^-\right \} \right ],\\[0.5em]
{\bf{v}}_4^{\circ}&=\frac{1}{M}\left [ m_1{\bf{v}}_1+m_2{\bf{v}}_2-
                           m_3\sqrt\frac{\mu_{12}}{\mu_{34}}\left \{ ({\bf{v}}_1-{\bf{v}}_2)-
                           \epsilon\left \langle \epsilon,({\bf{v}}_1-{\bf{v}}_2) \right \rangle +
                           \epsilon \omega ^-\right \} \right ],
\end{aligned}
\label{eq:PCVFR}
\end{equation}
where $\omega ^-=\sqrt{(\left \langle \epsilon,({\bf{v}}_1-{\bf{v}}_2) \right \rangle)^2-2Q_R/\mu_{12}}$. 
Analogously, the post-collisional velocities for the reverse chemical reaction 
$A_3+A_4 \rightarrow A_1+A_2$ 
are given by
\begin{equation}
\begin{aligned}
{\bf{v}}_1^{\circ}&=\frac{1}{M}\left [ m_3{\bf{v}}_3+m_4{\bf{v}}_4 +
        m_2\sqrt\frac{\mu_{34}}{\mu_{12}}\left \{ ({\bf{v}}_3-{\bf{v}}_4)
        -\epsilon\left \langle \epsilon,({\bf{v}}_3-{\bf{v}}_4) \right \rangle  +
        \epsilon \omega ^+ \right \} \right ] ,
        \\[0.5em]
{\bf{v}}_2^{\circ}&=\frac{1}{M}\left [ m_3{\bf{v}}_3+m_4{\bf{v}}_4-
        m_1\sqrt\frac{\mu_{34}}{\mu_{12}}\left \{ ({\bf{v}}_3-{\bf{v}}_4)
        -\epsilon\left \langle \epsilon,({\bf{v}}_3-{\bf{v}}_4) \right \rangle +
        \epsilon \omega ^+ \right \} \right ] ,
\end{aligned}
\label{eq:velr}
\end{equation}
where $\omega ^+ = \sqrt{(\left \langle \epsilon,({\bf{v}}_3-{\bf{v}}_4) \right \rangle)^2 + 2Q_R/\mu_{34}}$.

\bigskip


\noindent
We close this subsection by recalling some properties
about the dynamics of the reactive collisions
that have been established in \cite{CarvalhoPhD13} and \cite{PS17}.

\medskip

\begin{propt}
             \label{lem:properties}
             For a reactive collision, the following properties hold
             \begin{align}
             &\frac{1}{2}\mu_{ij}({\bf{v}}_i-{\bf{v}}_j)^2=\frac{1}{2}\mu_{kl}({\bf{v}}_k^{\circ}-{\bf{v}}_l^{\circ})^2 +Q_R ,
             \label{eq:activationenergy}\\
             &\mu_{ij}\left ( \left \langle\epsilon,{\bf{v}}_i-{\bf{v}}_j \right \rangle \right )^2
             =\mu_{kl}\left ( \left \langle \epsilon, {\bf{v}}_k^{\circ}-{\bf{v}}_l^{\circ} \right \rangle \right )^2+2Q_R,
             \label{eq:activationenergy1}\\
              &\frac{1}{2}\mu_{ij}\left ( \left \langle\epsilon,{\bf{v}}_i-{\bf{v}}_j \right \rangle \right )^2-\zeta_i
              =\frac{1}{2}\mu_{kl}\left ( \left \langle \epsilon, {\bf{v}}_k^{\circ}-{\bf{v}}_l^{\circ} \right \rangle \right )^2-\zeta_k,
              \label{eq:activationenergy2}\\
              & \left \langle\epsilon,{\bf{v}}_i-{\bf{v}}_j \right \rangle 
              =\left( \frac{\mu_{kl}}{\mu_{ij}} \right)^{\!\!\frac{1}{2}}\omega^+.
              \label{eq:activationenergy3}
             \end{align}
\end{propt}
\begin{propt}
\label{lem:Transformation}
For a fixed vector $\epsilon$, the Jacobians of the transformations 
$({\bf{v}}_i,{\bf{v}}_j) \mapsto ({\bf{v}}_k^{\circ},{\bf{v}}_l^{\circ})$ and 
$({\bf{v}}_k,{\bf{v}}_l)\mapsto({\bf{v}}_i^{\circ},{\bf{v}}_j^{\circ})$
are, respectively, given by
\begin{equation}
\left( \frac{\mu_{kl}}{\mu_{ij}} \right)^{\!\!\frac{3}{2}}
       \frac{\left \langle \epsilon,{\bf{v}}_k^{\circ}-{\bf{v}}_l^{\circ} \right \rangle}{\omega^+}
       \quad \text{and} \quad  
\left( \frac{\mu_{ij}}{\mu_{kl}} \right)^{\!\!\frac{3}{2}}
       \frac{\left \langle \epsilon,{\bf{v}}_i-{\bf{v}}_j \right \rangle}{\omega^-}.
\label{eq:Jacobian_transformation_vivjvkvl}
\end{equation}
\end{propt}

\medskip


\subsection{Kinetic equations}
\label{ssec:keqs}

The state of the reactive mixture is described by the one-particle distribution functions 
$f_i(t,{\bf{x}},{\bf{v}}_i)$ representing the density of particles of species $A_i$, 
expressed in moles, 
which at time $t$ are located at position ${\bf{x}}$ and have velocity ${\bf{v}}_i$, 
with $i=1,2,3,4$ and $(t, {\bf{x}}, {\bf{v}}_i) \in \mathbb{R}_+ \times \Omega \times \mathbb{R}^3$.  
The functions $f_i$ are related to the molar concentrations $c_i$ through the following expressions
\begin{equation}
c_i(t,{\bf{x}})=\int_{\mathbb{R}^3}f_i(t,{\bf{x}},{\bf{v}}_i)\,d{\bf{v}}_i, 
\qquad  t\geq 0, \;\; {\bf{x}}\in\Omega, \;\; i=1,2,3,4 .
\label{eq:C_I}
\end{equation}
In absence of external forces, the SRS kinetic equations are given by
\begin{equation}
\frac{\partial f_i}{\partial t}+{\bf{v}}_i\cdot \frac{\partial f_i}{\partial {\bf{x}}} = J_i \, ,
             \quad \text{in}\quad \mathbb{R}_+ \times \Omega \times \mathbb{R}^3,
\label{eq:KE}
\end{equation}             
with $J_i=J_i^E + J_i^R$, for $ i=1,2,3,4$,
where $J_i^E$ is the elastic collision operator and $J_i^R$ is the reactive collision operator. 
They are respectively defined as follows,
\begin{equation}
\begin{aligned}
J_i^E&=  \sigma_{ii}^2 \int _{{\mathbb{R}}^3}\int _{{\mathbb{S}^2_+}}
             \left [ {f}_i{'}{f'_{i_*}}-f_if_{i_*} \right ]\left \langle \epsilon ,{\bf{v}}_i-{\bf{v}}_{i_*} \right \rangle d\epsilon \,d{\bf{v}}_{i_*} \\                
              &\qquad+ \sum_{\substack{s=1\\ s\neq i}}^{4}
              \sigma^2 _{is} \int_{{\mathbb{R}}^3} \int _{{\mathbb{S}^2_+}}
              \left [ {f}_i{'}{f'_s}-f_if_s \right] \left \langle \epsilon ,{\bf{v}}_i-{\bf{v}}_s \right\rangle d\epsilon\, d{\bf{v}}_s \\
              &\qquad\qquad-\beta_{ij}\sigma^2_{ij} 
              \int_{{\mathbb{R}}^3}\int_{{\mathbb{S}^2_+}}
              \left [ {f_i}{'}{f_j}{'}-f_if_j \right ]\Theta \left( \left\langle \epsilon ,{\bf{v}}_i-{\bf{v}}_j \right\rangle-\Xi _{ij} \right)
              \left\langle \epsilon , {\bf{v}}_i-{\bf{v}}_j \right\rangle d\epsilon \, d{\bf{v}}_j ,
\end{aligned}
\label{eq:JiE}
\end{equation}
\begin{equation}
J_i^{R}  = \beta_{ij}\sigma^2 _{ij} \! \int_{{\mathbb{R}}^3} \int_{{\mathbb{S}^2_+}} \!
              \left[ \!\left ( \! \frac{\mu_{ij}}{\mu_{kl}} \! \right )^{\!\!2} \!\! f_k^\circ f_l^\circ -f_if_j \right] \!
              \Theta \left ( \left \langle  \epsilon ,{\bf{v}}_i-{\bf{v}}_j \right \rangle-\Xi _{ij} \right )
              \left \langle \epsilon ,{\bf{v}}_i-{\bf{v}}_j  \right \rangle d\epsilon \,d{\bf{v}}_j ,
\label{eq:JiR}
\end{equation}
where we have adopted the usual notation
${f}_i' \!=\! f(t,{\bf{x}},{\bf{v}}'_i)$, 
$ {f}_{i*}' \!=\! f(t,{\bf{x}},{\bf{v}}'_{i*})$, 
${f}_s' \!=\! f(t,{\bf{x}},{\bf{v}}'_s)$, 
${f_k}^{\circ} \!=\! f(t,{\bf{x}},{\bf{v}}_k^{\circ})$, 
${f_l}^{\circ} \!=\! f(t,{\bf{x}},{\bf{v}}_l^{\circ})$,
and
$\Theta$ is a Heaviside step function, defined at $x\in\mathbb {R}$ by
\begin{equation}
\Theta (x) 
= \left\{
             \begin{array}{ll}
             1, & \quad x \geq 0  ,\\
             0, & \quad x < 0. 
             \end{array} \right.
\label{eq:indic}
\end{equation}
Equations \eqref{eq:KE} together with expressions  \eqref{eq:JiE} and \eqref{eq:JiR}
constitute the SRS kinetic system. 
Without being precise, the accompanying  boundary conditions 
to describe the interactions between the molecules and the boundary $\partial\Omega $ 
of the evolution domain are taken to be of specular reflection type 
\cite{Villani}.   
Such boundary conditions ensure that the reactive mixture is considered in a closed domain,
as assumed in Section \ref{sec:ms}.

\medskip

\noindent
In expression \eqref{eq:JiE} for the elastic collision operator,  
the first term on the right hand side represents collisions involving particles of the same species 
and the index $i_*$ is used to distinguish their velocities.
Such term represents the standard Boltzmann collision operator for a single gas (mono-species)
and will be denoted by $J_i^{mE}$.
The second term in the same expression describes elastic collisions between particles of different species
(bi-species)
and will be denoted by $J_i^{bE}$.
The third term singles out the fraction $\beta_{ij}$ of those pre-collisional states 
that are energetic enough to result in chemical reaction,
and thus prevent double counting of these collisions in the elastic and reactive operators.
Such term will be denoted by $J_{ij}^{b*E}$.
Accordingly, in what follows, for $i=1,2,3,4$, we will write
\begin{equation}
J_i^E = J_i^{mE} + J_i^{bE}-J_{ij}^{b*E} ,
\label{eq:JiEmb}
\end{equation}
with the following notations
\begin{align}
& J_i^{mE} =\sigma_{ii}^2 \int _{{\mathbb{R}}^3}\int _{{\mathbb{S}^2_+}}
               \left [ {f}_i{'}{f'_{i_*}}-f_if_{i_*} \right ]
               \left \langle \epsilon ,{\bf{v}}_i-{\bf{v}}_{i_*} \right \rangle d\epsilon \,d{\bf{v}}_{i_*} ,
               \label{eq:Jime} \\
& J_i^{bE} = \sum_{\substack{s=1\\ s\neq i}}^{4}
               \underbrace{\sigma^2 _{is} \int _{{\mathbb{R}}^3}
               \int_{{\mathbb{S}^2_+}}\left [ {f}_i^{'}{f_s}^{'}-f_if_s \right ]
               \left \langle \epsilon ,{\bf{v}}_i-{\bf{v}}_s \right \rangle d\epsilon  d{\bf{v}}_s}_{{\cal Q}_{is}} ,
               \label{eq:Jibe} \\
& J_{ij}^{b*E}   
               = \beta_{ij}\sigma^2_{ij} \int _{{\mathbb{R}}^3}
               \int_{{\mathbb{S}^2_+}} \left [ {f_i}^{'}{f_j}^{'}-f_if_j \right ]
               \Theta \left ( \left \langle \epsilon ,{\bf{v}}_i-{\bf{v}}_j \right \rangle -\Xi _{ij} \right )
               \left \langle \epsilon ,{\bf{v}}_i-{\bf{v}}_j \right \rangle d\epsilon \, d{\bf{v}}_j .
               \label{eq:jib*e}
\end{align}

\noindent
The following proposition provides an alternative form for the collision operator $J_i$,
which is very useful to interpret the collisional dynamics of the model and, in particular, 
the role of the operator $J_{ij}^{b*E}$. 
For the proof, it is enough to combine the two contributions of the elastic operator 
relative to species $i$ and $j$ and use the identity $\Theta(x)+\Theta(-x)=1$.

\medskip

\begin{prop}
For each $i=1,2,3,4$, the collision operator $J_i$ introduced in \eqref{eq:KE} can be written as
\begin{align}
J_i
             &= \sigma_{ii}^2 \int _{{\mathbb{R}}^3}\int _{{\mathbb{S}^2_+}}
             \left [ {f}_i'{f'_{i_*}}-f_if_{i_*} \right ]
             \left \langle \epsilon ,{\bf{v}}_i-{\bf{v}}_{i_*} \right \rangle d\epsilon \,d{\bf{v}}_{i_*}
             \nonumber  \\   
             &+\sigma^2 _{ik} \int _{{\mathbb{R}}^3}\int_{{\mathbb{S}^2_+}} \!
             \left [ {f}_i{'}f_k' \!-\! f_if_k \right ] \!
             \left \langle \epsilon ,{\bf{v}}_i  \!-\!  {\bf{v}}_k \right \rangle d\epsilon  d{\bf{v}}_k
             +\sigma^2 _{il} \int _{{\mathbb{R}}^3}\int_{{\mathbb{S}^2_+}} \!
             \left [ {f}_i'f_l' \!-\! f_if_l \right ] \!
             \left \langle \epsilon ,{\bf{v}}_i \!-\! {\bf{v}}_l \right \rangle d\epsilon  d{\bf{v}}_l 
             \nonumber \\         
             &+ \sigma^2 _{ij} \int _{{\mathbb{R}}^3}\int_{{\mathbb{S}^2_+}}
             \Theta \left (\Xi _{ij} - \left \langle \epsilon ,{\bf{v}}_i-{\bf{v}}_j \right \rangle \right )
             \left [ {f}_i{'}{f_j}{'}-f_if_j \right ]
             \left \langle \epsilon ,{\bf{v}}_i-{\bf{v}}_j \right \rangle d\epsilon  d{\bf{v}}_j
             \label{eq:JiEplusJiR}  \\
             &+ (1-\beta_{ij}) \sigma^2 _{ij} \int _{{\mathbb{R}}^3}\int_{{\mathbb{S}^2_+}}
             \Theta \left ( \left \langle \epsilon ,{\bf{v}}_i-{\bf{v}}_j \right \rangle -\Xi _{ij} \right ) 
             \left [ {f}_i{'}{f_j}{'}-f_if_j \right ]
             \left \langle \epsilon ,{\bf{v}}_i-{\bf{v}}_j \right \rangle d\epsilon  d{\bf{v}}_j
             \nonumber  \\
             &+\beta_{ij}\sigma^2 _{ij} \! \int_{{\mathbb{R}}^3} \int_{{\mathbb{S}^2_+}} \!
             \left[ \!\left ( \!\frac{\mu_{ij}}{\mu_{kl}} \! \right )^{\!\!2} \!\! f_k^\circ f_l^\circ -f_if_j \right] \!
             \Theta \left ( \left \langle  \epsilon ,{\bf{v}}_i-{\bf{v}}_j \right \rangle-\Xi _{ij} \right )
             \left \langle \epsilon ,{\bf{v}}_i-{\bf{v}}_j \right \rangle d\epsilon \,d{\bf{v}}_j .
             \nonumber
\end{align}
\label{propnnn}
\end{prop}


\medskip

\noindent
Let us focus on the last three integrals on the right hand side of expression \eqref{eq:JiEplusJiR}, 
which are associated to collisions between the reactive species $A_i$ and $A_j$. 
The first of these integrals, with $\sigma^2 _{ij}$ in front of it,
is related to those collisions between $A_i, A_j$ with insufficient amount of energy to produce a chemical reaction, 
and therefore are governed by elastic collisional dynamics.
The last two integrals correspond to collisions between $A_i, A_j$ with sufficient amount of energy to produce a chemical reaction. 
However, only a fraction $\beta_{ij}$ of such collisions results in a chemical reaction (last integral)
and produces species $A_k$ and $A_l$.
The remaining fraction $(1-\beta_{ij})$ corresponds to collisions that
are also governed by elastic dynamics (second to the last integral).

\medskip

\begin{rem}
Observe that by setting the coefficients $\beta_{ij}$ equal to zero, 
the collisional terms $J_i^{b*E}$ and $J_i^R$ vanish,
see  \eqref{eq:JiR} and \eqref{eq:jib*e}.
This corresponds to a situation in which the chemical reaction is turned off
and we recover from our equations the hard-spheres model for a non-reactive mixture.
Moreover, by setting the coefficients equal to one,  
all collisions with sufficient amount of energy to produce a chemical reaction
will result, in fact, in a reactive collision.
However, this is not the case in general, because it is well known in chemistry \cite{Zum12}
that besides the activation energy barrier, the relative orientation of the molecules at the instant of collision is very important 
for the occurrence of a chemical reaction,
meaning that only collisions with sufficient amount of energy 
and right orientation will result in a chemical reaction.
Accordingly, we will consider in this paper the case in which $\beta_{ij} \in \; ]0,1[$
to guarantee that chemical reaction in fact occurs ($\beta_{ij} > 0$)
but some collisions between the reactive species will not result in a chemical reaction 
due to improper orientation ($\beta_{ij} < 1$),
even if they have enough energy to react chemically.           
\label{rmknnn}
\end{rem}

\medskip


\subsection{Fundamental Properties of The SRS Model}
\label{ssec:fp}

In this subsection we review some fundamental properties of the SRS kinetic system.
We have decided to include these properties in our paper because 
here we split the elastic collision operator in a particular form, 
see \eqref{eq:JiEmb}. 
These properties are adapted to our formalism and can be proved 
adapting the proofs in \cite{CarvalhoPhD13, GP04, Polewczak00, PS17} for similar results.

    
\medskip

\begin{lem}
\label{lem:weakform_MS}
Given the mono-species elastic collision operator $J_i^{mE}$, 
let $\varphi({\bf{v}}_i)$ be a sufficiently smooth test function.
Then, the weak form of  \eqref{eq:Jime} for each of the species in the reactive mixture is given by
\begin{multline}
\int_{\mathbb{R}^3}J_i^{mE}\varphi({\bf{v}}_i)\,d{\bf{v}}_i 
      =  \frac{1}{4} \sigma_{ii}^2 \int_{\mathbb{R}^3}\int _{{\mathbb{R}}^3}\int _{{\mathbb{S}^2_+}}
      \left [ {f}_i^{'}{f}_{i*}^{'}-f_if_{i*} \right ]\left \langle \epsilon ,{\bf{v}}_i-{\bf{v}}_{i*} \right \rangle 
      \\
      \times\left [ \varphi({\bf{v}}_i)+\varphi({\bf{v}}_{i*})-\varphi({\bf{v}}'_i)-\varphi({\bf{v}}'_{i*}) \right ] 
      d\epsilon \,d{\bf{v}}_{i*} d{\bf{v}}_i.
\label{eq:weakform_MSECO}
\end{multline}
\end{lem}

\medskip

\begin{lem}
\label{lem:weakform_BS}
Given the bi-species elastic collision operator $J_i^{bE}$
defined in \eqref{eq:Jibe} as a sum of several contributions ${\cal Q}_{is}$,
let $\varphi({\bf{v}}_i)$ be a sufficiently smooth test function.
Then, for each $i,s=1,2,3,4$ with $i\not=s$, 
we have that
\begin{equation}
\int_{\mathbb{R}^3}{\cal Q}_{is}\,\varphi({\bf{v}}_i)d{\bf{v}}_i
       = \sigma^2_{is}\int_{\mathbb{R}^3} \int_{\mathbb{R}^3} \int_{\mathbb{S}_+^2} 
       \left [ \varphi({\bf{v}}_i')-\varphi ({\bf{v}}_i) \right ] 
       f_if_s \left \langle \epsilon ,{\bf{v}}_i-{\bf{v}}_s \right \rangle 
       d\epsilon\, d{\bf{v}}_s\,d{\bf{v}}_i .
\label{eq:weakform_Qisdvi}
\end{equation}
\end{lem}

\medskip

\begin{lem}
\label{lem:weakform_BSC}
Given the elastic collision operator $J_{ij}^{b*E}$ defined in \eqref{eq:jib*e},
let  $\varphi({{\bf{v}}}_i)$ be a sufficiently smooth test function. 
If we assume that $\beta_{ij}=\beta_{ji}$, then 
for $ (i,j) \in \left \{ (1,2),(2,1),(3,4),(4,3) \right \}$, 
we have that
\begin{multline}
\int_{\mathbb{R}^3}J_{ij}^{b*E}\varphi({\bf{v}}_i)\,d{\bf{v}}_i 
     =\beta_{ij}\sigma^2_{ij} \int _{{\mathbb{R}}^3}\int_{\mathbb{R}^3}\int _{{\mathbb{S}^2_+}}
     \left [ \varphi({\bf{v}}_i') -\varphi({\bf{v}}_i) \right ]f_if_j\,\Theta 
     \left ( \left \langle \epsilon ,{\bf{v}}_i-{\bf{v}}_j \right \rangle-\Xi _{ij} \right )\\
     \times\left \langle \epsilon ,{\bf{v}}_i-{\bf{v}}_j \right \rangle d\epsilon \, d{\bf{v}}_j\,d{\bf{v}}_i ,
\label{eq:weakform_Cijdvi}
\end{multline}
where ${\bf{v}}_i'$ and ${\bf{v}}_j'$ are post-collisional velocities 
of the elastic encounters defined as in \eqref{eq:EPCV}
with the index $s$ replaced by $j$.
\end{lem}

\bigskip


\noindent
Concerning the reactive collision operator $J_i^R$, 
we have the following property.

\medskip

\begin{lem}
\label{lem:weakform_JiR}
Given the reactive collision operator $J_i^R$ defined in \eqref{eq:JiR}, let     
$\varphi({{\bf{v}}}_i)$ be a sufficiently smooth test function.
If we assume that $ \beta_{ij}=\beta_{ji}$
and $\beta_{12}\sigma^2_{12}=\beta_{34}\sigma^2_{34}$, then we have that
\begin{align}
             \sum_{i=1}^{4}\int_{\mathbb{R}^3}J_i^R\varphi_i({\bf{v}}_i)d{\bf{v}}_i
             & =  \beta_{12} \sigma_{12}^2 \! \int_{\mathbb{R}^3} \! \int_{\mathbb{R}^3} \!\int_{\mathbb{S}_+^2} \left [ \varphi_1+\varphi_2 - \varphi_3^\circ - \varphi_4^\circ \right ]\left [ \! \left ( \frac{\mu_{12}}{\mu_{34}} \right )^{\!\!2} \! f_3^{\circ}f_4^{\circ}-f_1f_2 \right ]
             \hspace*{0.5cm}
             \nonumber  \\[2mm]
             & \qquad \qquad  \times \Theta \left ( \left \langle \epsilon ,{\bf{v}}_1-{\bf{v}}_2 \right \rangle
             - \Xi_{12} \right )\left \langle \epsilon , {\bf{v}}_1-{\bf{v}}_2 \right \rangle d\epsilon \,d{\bf{v}}_2\,d{\bf{v}}_1
             \label{eq:weakform_JiR12}
             \\[3mm]
             & = \beta_{34}\sigma_{34}^2 \! \int_{\mathbb{R}^3} \!\int_{\mathbb{R}^3} \! \int_{\mathbb{S}_+^2} \!
             \left [ \varphi_3+\varphi_4-\varphi _1^\circ-\varphi_2^\circ \right ]\left[ \left ( \frac{\mu_{34}}{\mu_{12}} \right )^{\!\!2} \! f_1^\circ f_2^\circ - f_3 f_4 \right]
             \nonumber \\[2mm]
             & \qquad \qquad  \times  \Theta \left ( \left \langle \epsilon ,{\bf{v}}_3-{\bf{v}}_4 \right \rangle-\Xi_{34} \right )\left \langle \epsilon ,{\bf{v}}_3-{\bf{v}}_4 \right \rangle  d\epsilon\, d{\bf{v}}_4\,d{\bf{v}}_3.
            \nonumber
            \end{align}
\label{eq:weakform_JiR34}
\end{lem}


\subsection{Conservation equations}
\label{ssec:cl}

The conservation equations of the SRS model
are obtained from the properties stated in 
Subsection \ref{ssec:fp}.
Their proofs are rather standard
and follow the same line as in \cite{CarvalhoPhD13, GP04, Polewczak00, PS17}.

\medskip

\begin{cor} 
              \label{cor:monospecies_MMKE_con}
             The  mono-species  elastic  collision  operator given in \eqref{eq:Jime} is such that, for $i=1,2,3,4$,
             \begin{equation}
             \int_{\mathbb{R}^3}J_i^{mE}
             \begin{pmatrix}1\\ 
              m_i{\bf{v}}_i\\ 
             \frac{1}{2}m_i({\bf{v}}_i)^2
             \end{pmatrix}d{\bf{v}}_i =0.
             \label{eq:Mconservation_MMKE}
             \end{equation}
\end{cor}

\noindent
{\it Proof.} The proof follows from Lemma \ref{lem:weakform_MS} 
and conservation laws \eqref{eq:ECOMAK}. 
\hfill$\square$

\medskip 

\begin{cor}
\label{cor:bispecies_MMKE_con}
Let ${\cal Q}_{is}$ and $J_{ij}^{b*E}$ be as defined in \eqref{eq:Jibe} and \eqref{eq:jib*e},
respectively. 
Then
             \begin{align}
              &\int_{\mathbb{R}^3}{\cal Q}_{is}\,d{\bf{v}}_i=0,
              \label{eq:Mconservation_BS}
              \\
              &\int_{\mathbb{R}^3}J_{ij}^{b*E}\,d{\bf{v}}_i=0,
              \label{eq:Mconservation_BSC}
              \\
              &\int_{\mathbb{R}^3}{\cal Q}_{is}
              \begin{pmatrix}
             m_i{\bf{v}}_i\\ 
             \frac{1}{2}m_i({\bf{v}}_i)^2
             \end{pmatrix}\,d{\bf{v}}_i+ \int_{\mathbb{R}^3}{\cal Q}_{si}
             \begin{pmatrix} 
             m_s{\bf{v}}_s\\ 
             \frac{1}{2}m_s({\bf{v}}_s)^2
             \end{pmatrix}\,d{\bf{v}}_s=0,
             \label{eq:MKEconservation_BS}
             \\
             &\int_{\mathbb{R}^3}J_{ij}^{b*E}
             \begin{pmatrix}
             m_i{\bf{v}}_i\\ 
             \frac{1}{2}m_i({\bf{v}}_i)^2
             \end{pmatrix}\,d{\bf{v}}_i+ \int_{\mathbb{R}^3}J_{ij}^{b*E}
             \begin{pmatrix} 
             m_j{\bf{v}}_j\\ 
             \frac{1}{2}m_j({\bf{v}}_j)^2
             \end{pmatrix}d{\bf{v}}_j=0.
             \label{eq:MKEconservation_BSC}
             \end{align}
\end{cor}

\noindent
{\it  Proof.} 
The proof follows from Lemmas \ref{lem:weakform_BS}, \ref{lem:weakform_BSC} 
and conservation laws \eqref{eq:ECOMAK}. 
\hfill$\square$

\medskip


\begin{cor}
The reactive collision operators satisfy the following property
             \begin{equation}
             \int_{\mathbb{R}^3}J^R_1 d{\bf{v}}_1  = \int_{\mathbb{R}^3}J^R_2d{\bf{v}}_2
             = - \int_{\mathbb{R}^3}J^R_3d{\bf{v}}_3 = - \int_{\mathbb{R}^3}J^R_4d{\bf{v}}_4.
             \label{eq:JiR1234}
             \end{equation}
\label{cr:JiR}
\end{cor}

\noindent
{\it  Proof.} The proof follows from Lemma \ref{lem:weakform_JiR}. 
\hfill$\square$

\medskip


\noindent
Corollary \ref{cr:JiR} assures the correct exchange rates 
for the species in the chemical reaction \eqref{eq:reac}.


\medskip

\begin{cor}
The elastic and reactive collision operators are such that
             \begin{equation}
             \sum_{i=1}^4 \int_{\mathbb{R}^3} \varphi({\bf{v}}_i) \Big( J^E_i + J^R_i \Big) d{\bf{v}}_i = 0,
             \label{eq:JiER}
             \end{equation}
with $\varphi ({\bf{v}}_i)$
alternatively given by
             $\varphi ({\bf{v}}_i) \!=\! \big( 1,0,1,0 \big)$, $\varphi ({\bf{v}}_i) \!=\! \big( 1,0,0,1 \big)$,
             $\varphi ({\bf{v}}_i) \!=\! \big( 0,1,1,0 \big)$, 
or by $\varphi({\bf{v}}_i) \!=\! m_i v_{ix}$,
             $\varphi({\bf{v}}_i) \!=\! m_i v_{iy}$, $\varphi({\bf{v}}_i) \!=\! m_i v_{iz}$, 
or by 
$\varphi({\bf{v}}_i) \!=\! E_i + \frac12 m_i {\bf v}_i^2$, where $v_{ix}$, $v_{iy}$ and $v_{iz}$ represent the spatial components of the molecular velocity ${\bf v}_i$. 
\label{pr:cons}
\end{cor}

\medskip


\noindent
{\it  Proof.} 
The proof follows from Corollaries \ref{cor:monospecies_MMKE_con}, \ref{cor:bispecies_MMKE_con} and
Lemma \ref{lem:weakform_JiR}. 
\hfill$\square$

\medskip


\medskip

\noindent
Corollary \ref{pr:cons} indicates that, at least formally, 
the SRS model possesses seven independent macroscopic conservation laws,
for the total number of particles of the reactant-product pairs of the form 
$A_1$-$A_3$, $A_1$-$A_4$ and $A_2$-$A_3$,
the three momentum components and the total energy of the mixture.

\medskip


\subsection{Equilibrium solutions and H-Theorem}
\label{ssec:esht}

The equilibrium solutions of the SRS system are characterized as follows.

\medskip

\begin{defn}
The equilibrium solutions of the SRS model \eqref{eq:KE} are distribution functions $f_i(t,{\bf x}, {\bf v} )$ such that
the operators $J_{i}^E$ and $J_{i}^R$ given in
\eqref{eq:JiE} and \eqref{eq:JiR} satisfy
\begin{equation}
J_{i}^E  + J_{i}^R  = 0, \qquad  i = 1, \ldots , 4.
 \label{theq}
\end{equation}
\end{defn}


\begin{prop} 
If the coefficients $\beta_{ij}$ and the reactive cross sections are such that
             $\beta_{ij}=\beta_{ji}$ and 
             $\beta_{12}\sigma^2_{12}=\beta_{34}\sigma^2_{34}$, 
             then the following statements are equivalent

\smallskip

\noindent
(a) $f_i = n_i \left(\dfrac{m_i}{2\pi kT}\right)^{\!3/2}\exp{\left(-\dfrac{m_i ({\bf v}_i - {\bf u})^2}{2kT}\right)}$, \ for \ $i=1,\dots,4$, \ with
             \begin{equation}
             n_1n_2 = \left( \frac{\mu_{12}}{\mu_{34}} \right)^{\!\!1/2} \! n_3n_4 \; \exp\left( \frac{Q_R}{kT}\right) ;
              \label{eq:mal}
              \end{equation}

\smallskip

\noindent
(b) $J_i^E =0$ \ and \ $J_i^R =0$, \ $i=1,\dots,4$;

\smallskip
          
\noindent
(c) $\dps \sum\limits_{i=1}^4 \, \int\limits_{\mathbb{R}^3}\left[J_i^E + J_i^R \right]\log\left(f_i/\mu_{ij}\right)\,dv=0$,

\smallskip

\noindent
where
             \begin{subequations}
             \begin{align}
             & n_i (t,{\bf x})  =  \int_{\mathbb{R}^3} f_i(t,{\bf x},{\bf v}_i) d{\bf v}_i, \qquad  i = 1, \ldots , 4,   
             \label{subeq:a} \\
             & {\bf u} (t,{\bf x})     =  \sum_{i=1}^4 \int_{\mathbb{R}^3} m_i {\bf v}_i f_i(t,{\bf x},{\bf v}_i) d{\bf v}_i\bigg/  
             \sum_{i=1}^4 \int_{\mathbb{R}^3} m_i f_i(t,{\bf x},{\bf v}_i) d{\bf v}_i,  
             \label{subeq:b} \\
             & T (t,{\bf x})     =  \frac1{3k} \; \sum_{i=1}^4 \int_{\mathbb{R}^3} m_i ({\bf v}_i-{\bf u})^2 
             f_i(t,{\bf x},{\bf v}_i) d{\bf v}_i  \bigg/  \sum_{i=1}^4 n_i (t,{\bf x}) .
             \label{subeq:c} 
             \end{align}
             \label{nuT}
             \end{subequations}
             \label{pr:equil}
             \end{prop}
             \medskip\\
 Condition \eqref{eq:mal}  
 represents the so called {\it mass action law} for the SRS kinetic model.          


\section{Reaction diffusion limit of the SRS kinetic model}
\label{sec:rdl}

In this section we formally derive the reaction-diffusion system of MS type    
as a hydrodynamic limit of the SRS kinetic model given in Section \ref{sec:srs}. 
In order to achieve this, we have to define an evolution regime for the chemical process 
and consider the mathematical assumptions that have to be imposed to the kinetic model 
in agreement with the physical conditions associated to the MS setting.

\subsection{\bf The scaled equations and our assumptions}

The starting point for the derivation of the reaction diffusion system of MS type
is the scaled SRS kinetic system in a form compatible with the considered chemical regime
of dominant elastic collisions and slow chemical reactions.
Accordingly, we will consider the SRS kinetic equations written in non-dimensional form as
\begin{equation}
St\frac{\partial \widehat{f_i}}{\partial \widehat{t}}+
           {\bf{\widehat{v}}}_i\cdot\frac{\partial \widehat{f_i}}{\partial{\bf{\widehat{x}}}}=
           \frac{1}{Kn_e}{\widehat{J}_i^{mE}}+\frac{1}{Kn_e}{\widehat{J}_i^{bE}}-
           \frac{1}{Kn_r}{\widehat{J}_i^{b*E}}+\frac{1}{Kn_r}\widehat{J}_i^R ,
\label{eq:dimensionless_KE2}
\end{equation}
where $St$ is the kinetic Strouhal number \cite{Golse12}  
and
$Kn_e$ and $Kn_r$ are the elastic and reactive Knudsen numbers \cite{ACG94}.
The Knudsen numbers are such that the reactive and elastic mean free paths are related by the factor 
$Kn_r / Kn_e = 1 / \beta_{ij}$.
The symbols with the hat indicate scaled quantities 
with respect to a reference length $L$, time $\tau$ and temperature $T_0$.
We also introduce the speed of sound $c_0$ in a mixture of
monatomic ideal gases at temperature $T_0$,
given by
$c_0=\sqrt{\frac{5n_0 k_B T_0}{3\rho_0}}$,
and scale the velocities with respect to $c_0$.
The scaled collisional operators are defined by
\begin{align*}
&{\widehat{J}_i^{mE}}= \widehat{\sigma}_{ii}\int _{\mathbb{R}^3} \int_{\mathbb{S}_+^2}\left [ \widehat{f_i}'\widehat{f}_{i*}'-\widehat{f_i}\widehat{f}_{i*} \right ]\left \langle \epsilon ,{\bf{\widehat{v}}}_i-{\bf{\widehat{v}}}_{i*} \right \rangle\, d\epsilon \,d{\bf{\widehat{v}}}_{i*},
\\
& {\widehat{J}_i^{bE}}=\sum_{\substack{s=1\\ s\neq i}}^{4} \widehat{\sigma} ^2_{is} \int _{\mathbb{R}^3} \int_{\mathbb{S}_+^2}\left [ \widehat{f_i}'\widehat{f_s}'-\widehat{f_i}\widehat{f_s} \right ]\left \langle \epsilon ,{\bf{\widehat{v}}}_i-{\bf{\widehat{v}}}_s \right \rangle\, d\epsilon\, d{\bf{\widehat{v}}}_s ,
\\
& {\widehat{J}_i^{b*E}}=\widehat{\sigma}_{ij}^2 
            \int_{\mathbb{R}^3} \int_{\mathbb{S}_+^2}\left [ \widehat{f_i}'\widehat{f_j}'-\widehat{f_i}\widehat{f_j} \right ]\Theta \left (  \left \langle \epsilon ,c_0{\bf{\widehat{v}}}_i-c_0{\bf{\widehat{v}}}_j \right \rangle-\Xi_{ij} \right )\left \langle  \epsilon ,{\bf{\widehat{v}}}_i-{\bf{\widehat{v}}}_j \right \rangle\,d\epsilon \,d{\bf{\widehat{v}}}_j,
\\
& {\widehat{J}_i^{R}}=\widehat{\sigma}_{ij}^2\int_{\mathbb{R}^3} \int_{\mathbb{S}_+^2}\left [\left( \frac{\mu_{ij}}{\mu_{kl}} \right)^{\!\!2} \widehat{f}^{\circ}_k \widehat{f}^{\circ}_l-\widehat{f_i}\widehat{f_j}\right ]\,\Theta \left (  \left \langle \epsilon ,c_0{\bf{\widehat{v}}}_i-c_0{\bf{\widehat{v}}}_j \right \rangle-\Xi_{ij} \right ) \left \langle  \epsilon ,{\bf{\widehat{v}}}_i-{\bf{\widehat{v}}}_j \right \rangle d\epsilon\, d{\bf{\widehat{v}}}_j.
\end{align*}
Moreover, we assume that 
the bulk velocity of the mixture is comparable to $L/\tau$ 
and, as a result, the Strouhal number becomes the Mach number, $Ma$.
Henceforth, we introduce the notations
\begin{equation}
St=Ma=\alpha, \qquad Kn_e=\alpha^p \qquad \text{and}\qquad Kn_r=\alpha^q,
\label{eq:StMa}
\end{equation}
where $\alpha \ll 1$, $p$ and  $q$ are real numbers with $p\geq1$ and $p>q$.
Also, we assume that the elastic Knudsen number is of the same order
of magnitude as the Mach number, so that $p=1$.
Additionally, we are interested in a chemical regime for which elastic collisions are predominant
and reactive collisions are rare, so that we assume $q=-1$.
Using the above notations and the generic assumptions in \eqref{eq:dimensionless_KE2},
and removing the hats for simplicity, 
we obtain
\begin{equation}
\alpha\frac{\partial {f_i^\alpha}}{\partial {t}} 
           + {\bf{{v}}}_i\cdot\frac{\partial {f_i^\alpha}}{\partial{\bf{{x}}}}
           = \frac{1}{\alpha}{{J}_i^{m\alpha E}}+\frac{1}{\alpha}{{J}_i^{b\alpha E}}
           -  \alpha{{J}_i^{b*\alpha E}} + \alpha{J}_i^{\alpha R} ,
           \quad \text{in} \quad 
           {\mathbb{R}_+}\times\Omega\times{\mathbb{R}^3} ,
\label{eq:scaled_KE}
\end{equation}
where $f_i^\alpha$, $i=1,2,3,4$, are now the scaled unknowns in the considered regime.
Equations \eqref{eq:scaled_KE} emphasize that the
elastic operators $J_i^{m\alpha E}$, $J_i^{b\alpha E}$ are associated to predominant effects, 
whereas the reactive operator $J_i^{m\alpha R}$ and the correction elastic term $J_i^{b*\alpha E}$ 
are associated to rare effects. 

\medskip

\noindent
With the obvious adjustments, the conservation laws and properties of the SRS model given in Section \ref{sec:srs} 
are still valid for the scaled equations \eqref{eq:scaled_KE}.


\medskip

\noindent
Besides the chemical regime of slow chemical reaction, 
our central assumptions to derive the reaction diffusion system of MS type 
from the kinetic formulation are as follows:

\begin{enumerate}[label={(\alph*)}]
\item \label{assM} The system evolves not far way from a local Maxwellian equilibrium state.
\item \label{assv} The bulk velocities of the species in the reactive mixture are small and go to zero as $\alpha \to 0$.
\item \label{assT} The evolution of the mixture obeys the isothermal condition. 
\end{enumerate}

\medskip

\noindent
In particular,
assumption \ref{assT} implies that the temperature $T$ of the mixture is uniform in space and constant in time.
%
From assumption \ref{assM}, the initial conditions for \eqref{eq:scaled_KE} are assumed to be local Maxwellian functions
centered at the average mean velocity of the species.
From the first assertion of \eqref{eq:StMa} and following \cite{Golse12,golsedaf},
such local Maxwellians are assumed to be of the form 
\begin{equation}
              f_i^{\alpha(in)} \! ({\bf{x}},{\bf{v}}_i) = c_i^{\alpha(in)} \! ({\bf{x}}) \!\!
              \left ( \! \frac{m_i}{2\pi k_BT} \! \right )^{\!\!\frac{3}{2}} \!
              \exp \! \left[-\frac{m_i \Big( \! {\bf{v}}_i \!-\! \alpha{\bf{u}}^{\alpha(in)}_i({\bf{x}}) \! \Big)^2}{2k_BT}\right] \! ,
              \;\; {\bf{x}}\in \Omega, \; {\bf{v}}_i\in \mathbb{R}^3,
\label{eq:local_max}
\end{equation}
where $T > 0$ is constant (isothermal conditions), 
$c_i^{\alpha(in)}  :  \, \Omega\rightarrow {\mathbb{R}}_+$, for $i=1, \ldots ,4$, with 
$c^{\alpha(in)} \!=\!  \sum_{i=1}^4 c_i^{\alpha(in)} $,
and ${\bf{u}}_i^{\alpha (in)} \! : \Omega\rightarrow {\mathbb{R}^3}$, for $i=1, \ldots ,4$.
Again, due to assumption \ref{assM}, the distribution functions $ f_i^{\alpha}(t,{\bf{x}},{\bf{v}}_i)$ 
remain local Maxwellians centered at the average mean velocity of the species,
\begin{equation}
f_i^{\alpha}(t,{\bf{x}},{\bf{v}}_i) =
       c_i^{\alpha}(t,{\bf{x}})
       \left ( \! \frac{m_i}{2\pi k_BT} \! \right )^{\!\!\frac{3}{2}} \!
       \exp \! \left[-\frac{m_i \Big( \! {\bf{v}}_i \!-\! \alpha{\bf{u}}^{\alpha}_i(t,{\bf{x}}) \! \Big)^2}{2k_BT}\right] \! ,
       \;\; {\bf{x}}\in \Omega, \; {\bf{v}}_i\in \mathbb{R}^3,
\label{eq:local_max1}
\end{equation}
where  $c_i^{\alpha}:{\mathbb{R}}_+ \times \Omega\rightarrow {\mathbb{R}}_+$ 
and $ {\bf{u}}_i^{\alpha}:{\mathbb{R}}_+ \times \Omega\rightarrow {\mathbb{R}^3}$,
for $i=1,\ldots4$.
Using \eqref{eq:local_max1}, the moments of order $0$ and $1$ 
of the distribution functions are, respectively, given by
\begin{align}
\label{eq:local_max_moments}
\int_{\mathbb{R}^3}f^\alpha_i(t,{\bf{x}},{\bf{v}}_i)d{\bf{v}}_i 
         = c^\alpha_i(t,{\bf{x}})
         \quad \mbox{and} \quad
\int_{\mathbb{R}^3}{\bf{v}}_if^\alpha_i(t,{\bf{x}},{\bf{v}}_i)d{\bf{v}}_i
         = \alpha c^\alpha_i(t,{\bf{x}}){\bf{u}}_i^\alpha(t,{\bf{x}}) .
\end{align}
An important step in the passage from a kinetic model to hydrodynamic equations
is the explicit computation of the integral contributions appearing in the kinetic equations.
When a reactive mixture is involved, 
such computation can be very intricate,
essentially because of the redistribution of masses and internal energies among the constituents.
Having this in mind, in the light of assumption \ref{assv}, 
the Maxwellian \eqref{eq:local_max1} with species velocity
will be approximated through a first order expansion around the Maxwellian 
$M_i^\alpha$ with a common and vanishing average mean velocity, 
as
\begin{equation}
f_i^{\alpha}(t,{\bf{x}},{\bf{v}}_i) 
       \approx 
       M_i^\alpha(t,{\bf{x}},{\bf{v}}_i) 
       \left[ 1 + \frac{\alpha m_i  {\bf{v}}_i {\bf{u}}_i^\alpha(t,{\bf x})}{k_BT} \right]  \! ,
\label{eq:local_maxapp}
\end{equation}
with
\begin{equation}
M_i^\alpha(t,{\bf{x}},{\bf{v}}_i)  = c_i^{\alpha}(t,{\bf{x}})
       \left ( \! \frac{m_i}{2\pi k_BT} \! \right )^{\!\!\frac{3}{2}} \!
       \exp \! \left( -\frac{m_i ({\bf{v}}_i )^2}{2k_BT} \right ) .
\label{eq:max0}
\end{equation}
Expressions \eqref{eq:local_maxapp} will be used in Subsections \ref{ssec:cbe} and \ref{ssec:mbe}
for the explicit computation of the integral contributions. 
As stated by Present in \cite{RDPresent58}, p.~147, 
first-order corrections to the Maxwellian distributions \eqref{eq:max0} 
as those given by expansions \eqref{eq:local_maxapp} 
were first used by J. Stefan in 
his celebrated work from 1872
to capture the diffusion effects in a mixture of gases.

Moreover, hereinafter we will use the notation $\Gamma(\eta,x)$
for the incomplete Euler gamma function defined by
\begin{equation}
\Gamma(\eta,x)=\int_x^\infty t^{\eta-1}e^{-t} dt .
\label{eq:int_incomplete_gamma_function}
\end{equation}


\subsection{Concentration balance equations}
\label{ssec:cbe}

The balance equations for the species concentration
can formally be derived from the scaled SRS equations given in \eqref{eq:scaled_KE},
by integrating over the velocity space. 
Also, the conservation laws  and some of the fundamental properties of the kinetic model stated in 
Section \ref{sec:srs} will be used to achieve the following result.

\medskip

\begin{lem}
\label{lem:con_balance}
The concentration balance equation for each of the species in the reactive mixture can be written as
\begin{equation}
\frac{\partial c_i^\alpha}{\partial t}+ \frac{\partial }{\partial {\bf{x}}}( c_i^\alpha {\bf{u}}_i^\alpha)
        = {\cal J}_i ,
        \qquad i=1,\ldots,4 ,
\label{eq:lcbe}
\end{equation}
with ${\cal J}_i$ being the reaction rate of the $i$-th species given by 
\begin{align}
{\cal J}_i = \sigma^2_{ij}\left( \frac{2\pi\mu_{ij}}{ k_BT} \right)^{\!\!\frac{1}{2}}
             & \left[ \left( \frac{\mu_{ij}}{\mu_{kl}} \right)^{\!\!\frac{1}{2}} \!
             c_k^\alpha\,c_l^\alpha\,
             \exp\left( {\frac{Q_R}{k_B T}}\right) - c_i^\alpha\,c_j^\alpha \right] 
             \label{eq:int_JialphaR}
             \\
             & \times \left[ \frac{2k_BT}{\mu_{ij}} \,
             \Gamma (2,z_i^*)-\left ( \frac{\Xi_{ij}}{c_0} \right )^{\!\!2} \; \Gamma (1,z_i^*) \right] \! ,
             \nonumber
\end{align}
where the terms $\Gamma (1,z_i^*), \, \Gamma (2,z_i^*)$ are defined by 
\eqref{eq:int_incomplete_gamma_function}
with 
\begin{equation}
z_i^* = \frac{\mu_{ij}}{2k_BT}\left ( \frac{\Xi_{ij}}{c_0} \right )^{\!\!\!2} .
\label{eq:zzii}
\end{equation}
Moreover, for each $i=1,\ldots,4$ we have 
${\cal J}_i=\lambda_i{\cal J}_1$, 
with $\lambda_1=\lambda_2 =1$ and $\lambda_3=\lambda_4 =-1$.
\end{lem}

\medskip

\noindent
{\it Proof.} 
First, we integrate both sides of the scaled SRS equations given in \eqref{eq:scaled_KE} 
with respect to ${\bf{v}}_i\in {\mathbb{R}^3}$, 
to obtain
\begin{equation*}
\alpha\frac{\partial }{\partial t} \left( \int_{\mathbb{R}^3}f^\alpha_i d{\bf{v}}_i \right)
       + \frac{\partial }{\partial {\bf{x}}} \left( \int_{\mathbb{R}^3}{\bf{v}}_if_i^\alpha d{\bf{v}}_i \right)
       = \alpha\int_{\mathbb{R}^3}J_i^{\alpha R} d{\bf{v}}_i ,
\end{equation*}
where we have used 
Corollary \ref{cor:monospecies_MMKE_con},
Lemma \ref{lem:weakform_BS} 
with $\varphi({\bf{v}}_i)=1$, 
and 
Lemma \ref{lem:weakform_BSC} 
with $\varphi({\bf{v}}_i)=1$. 
Using  \eqref{eq:local_max_moments},
we obtain 
\begin{equation}
\frac{\partial c_i^\alpha}{\partial t} + \frac{\partial }{\partial {\bf{x}}}( c_i^\alpha {\bf{u}}_i^\alpha)
         = \int_{\mathbb{R}^3}J_i^{\alpha R} d{\bf{v}}_i .
\label{eq:con_balance1}
\end{equation}
Next, we derive an explicit expression for the integral contribution on the right hand side of 
\eqref{eq:con_balance1},
replacing the distribution function by the approximation \eqref{eq:local_maxapp}.
For sake of simplicity, we write such approximations in the form
\begin{equation}
f_i^\alpha \approx M_i^\alpha(1+a_i{\bf{v}}_i) ,
    \qquad \mbox{with} \quad
    a_i=\frac{\alpha m_i{\bf{u}}_i^\alpha}{k_BT}.
\label{eq:local_maxapp1}
\end{equation}
Using \eqref{eq:local_maxapp1} 
and neglecting quadratic terms in the coefficients $a_i$, 
we can write the integral on the right-hand-side of \eqref{eq:con_balance1}
as the sum of the following three terms,
\begin{align}
&
{\cal D} \!=\! { \sigma^2 _{ij} \!\! \int _{{\mathbb{R}}^3} \! \int _{{\mathbb{R}}^3} \! \int _{{\mathbb{S}^2_+}}
          \!\!\!\! \left( \! \frac{\mu^2_{ij}}{\mu^2_{kl}} M_k^{\alpha\circ} \! M_l^{\alpha\circ} \!-\! M_i^\alpha \! M_j^\alpha \! \right)
          \!\! \Theta \! \left( \! \left \langle \epsilon ,{c_0\bf{v}}_i \!-\! {c_0\bf{v}}_j \right \rangle \!-\! \Xi _{ij} \right ) \!
          \left \langle \epsilon ,{\bf{v}}_i \!-\! {\bf{v}}_j \right \rangle d\epsilon d{\bf{v}}_j d{\bf{v}}_i} ,
          \nonumber \\[2mm]
& 
{\cal E} \!=\!  {\sigma^2 _{ij} \!\!\! \int _{{\mathbb{R}}^3} \! \int _{{\mathbb{R}}^3} \! \int _{{\mathbb{S}^2_+}} 
          \!\!\!\!\! \left( \!\! \frac{\mu^2_{ij}}{\mu^2_{kl}} M_k^{\alpha\circ} \! M_l^{\alpha\circ} \!
          (a_k \! {\bf{v}}_k^{\circ} \!+\! a_l \! {\bf{v}}_l^{\circ}) \!\!\!\! \right) 
          \!\!\! \Theta \! \left( \! \left \langle \epsilon ,{c_0\bf{v}}_i \!-\! {c_0\bf{v}}_j \right \rangle \!-\! \Xi _{ij} \right ) \!
          \left \langle \epsilon ,{\bf{v}}_i \!-\! {\bf{v}}_j \right \rangle d\epsilon d{\bf{v}}_j d{\bf{v}}_i} ,
          \nonumber \\[2mm]
&
{{\cal F}} \!=\!  - {\sigma^2 _{ij} \!\!\! \int _{{\mathbb{R}}^3} \! \int _{{\mathbb{R}}^3} \! \int _{{\mathbb{S}^2_+}}
          \!\!\!  \Big( \! M_i^\alpha M_j^\alpha (a_i{\bf{v}}_i \!+\! a_j{\bf{v}}_j) \! \Big)   
          \Theta \! \left( \! \left \langle \epsilon ,{c_0\bf{v}}_i \!-\! {c_0\bf{v}}_j \right \rangle \!-\! \Xi _{ij} \right ) \!
          \left \langle \epsilon ,{\bf{v}}_i \!-\! {\bf{v}}_j \right \rangle d\epsilon d{\bf{v}}_j d{\bf{v}}_i}.
          \label{eq:Jialpha_approx}
\end{align}
\noindent
Inserting \eqref{eq:max0} into the integral contributions
${\cal D}$, ${\cal E}$ and ${\cal F}$,
we obtain 

\medskip

(i) ${\cal D} = {\cal J}_i$, with ${\cal J}_i$ the reaction rate defined in \eqref{eq:int_JialphaR};

\medskip

(ii)  ${\cal E} = 0$;

\medskip

(iii) ${\cal F} = 0$.

\medskip

\noindent
Concerning items (ii) and (iii), the computations require some variable transformations
and we give some details in Appendix A of Subsection \ref{App:cbe}.
Let us focus on item (i).
Using the conservation law of total energy during reactive collisions
given in \eqref{eq:RCTE},
we obtain
\begin{align}
{\cal D} = 
        & \sigma^2 _{ij} \frac{(m_im_j)^{\!\!\frac{3}{2}} }{(2\pi k_BT)^3}
        \left[ \left( \frac{\mu_{ij}}{\mu_{kl}} \right)^{\!\!\frac{1}{2}} \!
        c_k^\alpha\,c_l^\alpha\,
        \exp\left( {\frac{Q_R}{k_B T}}\right) - c_i^\alpha\,c_j^\alpha \right]
        \label{eq:D} \\
        & \times \!\!
        \int _{{\mathbb{R}}^3} \! \int _{{\mathbb{R}}^3} \! \int _{{\mathbb{S}^2_+}} \!\!\!\!
        \exp\biggl[ - \frac{m_i ({\bf{v}}_i)^2 \!+\! m_j ({\bf{v}}_j)^2 }{2k_BT} \biggr] \!           
        \Theta \! \left ( \left \langle \epsilon ,{c_0\bf{v}}_i \!-\! {c_0\bf{v}}_j \right \rangle \!-\! \Xi _{ij} \right )
        \! \left \langle \epsilon ,{\bf{v}}_i-{\bf{v}}_j \right \rangle 
        d\epsilon d{\bf{v}}_j  d{\bf{v}}_i.
        \nonumber
\end{align}
Evaluating the integral over $\mathbb{S}_+^2$,    
using spherical coordinates as described in Remark \ref{rm:sphe},
we obtain 
\begin{equation}
\int_{\mathbb{S}_+^2} \!\! \Theta \left( \left\langle \epsilon,c_0{\bf{v}}_i \!-\! c_0{\bf{v}}_j \right\rangle \!-\! \Xi_{ij}\right)
          \left\langle \epsilon,{\bf{v}}_i-{\bf{v}}_j \right \rangle d\epsilon
       = \pi V \Theta \left( \! V \!-\! \frac{\Xi_{ij}}{c_0} \! \right ) \!\!
       \left [1 \!-\! \left( \frac{\Xi_{ij}}{c_0V} \right )^{\!\!2} \right ].
\label{eq:angular_int}
\end{equation}
Transforming the above sixfold integral in ${\bf v}_i , \, {\bf v}_j $
to the relative velocity $\bf V$ and centre of mass velocity ${\bf X}=(m_i {\bf v}_i + m_j {\bf v}_j)/(m_i+m_j)$, 
and using the fact that
the Jacobian of the transformation is equal to $1$,
we get
\begin{multline}
{\cal D} = 
        \pi \sigma^2 _{ij} \frac{(m_im_j)^{\!\!\frac{3}{2}} }{(2\pi k_BT)^3}
        \left[ \left( \frac{\mu_{ij}}{\mu_{kl}} \right)^{\!\!\!\frac{1}{2}} \!
        c_k^\alpha\,c_l^\alpha\,
        \exp\left( {\frac{Q_R}{k_B T}}\right) - c_i^\alpha\,c_j^\alpha \right]
        \label{eq:D_2}
        \\
        \times \!\!
        \int _{{\mathbb{R}}^3} \!\!
        \exp \! \left( \!\! - \frac{MX^2}{2k_B T} \right) \! d{\bf{X}}   
        \;\;
        \int _{{\mathbb{R}}^3} \!\! V \Theta \left( \! V \!-\! \frac{\Xi_{ij}}{c_0} \! \right ) \!\!
        \left [1 \!-\! \left( \frac{\Xi_{ij}}{c_0V} \right )^{\!\!2} \right ]
       \exp \! \left( \!\! - \frac{\mu_{ij}{{V}}^2}{2k_BT} \right) d{\bf{V}} , 
\end{multline}
where $M = m_i+m_j$.
The integral in $\bf X$ can be easily evaluated and becomes
\begin{align}
\int _{\mathbb{R}^3} \!\!
        \exp \! \left( \!\! - \frac{MX^2}{2k_B T} \right) \! d{\bf{X}}
        =
        \left( \frac{2\pi k_BT}{M} \right)^{\!\! \frac32}.
\label{eq:D_1_1}
\end{align}
The integral in $\bf V$ can be computed by first transforming to spherical coordinates.
Next, the resulting scalar integral in $V$ is transformed to
$z$, using $z= \frac{\mu_{ij}}{2k_BT} \, V^2$, and finally using \eqref{eq:int_incomplete_gamma_function}, 
we obtain
\begin{multline}
\int _{{\mathbb{R}}^3} \!\! V \Theta \left( \! V \!-\! \frac{\Xi_{ij}}{c_0} \! \right ) \!\!
        \left [1 \!-\! \left( \frac{\Xi_{ij}}{c_0V} \right )^{\!\!2} \right ]
        \exp \! \left( \!\! - \frac{\mu_{ij}{{V}}^2}{2k_BT} \right) d{\bf{V}}
        \label{eq:D_2n}
        \\
        = 
        2\pi \frac{2k_BT}{\mu_{ij}} 
        \left [ \frac{2k_BT}{\mu_{ij}} \; \Gamma (2,z^*) - \left( \frac{\Xi_{ij}}{c_0} \right)^{\!\!\!2} \;\Gamma (1,z^*)\right ] \! .
        \hspace*{1.5cm}
\end{multline}
Substituting \eqref{eq:D_1_1} and \eqref{eq:D_2n} into \eqref{eq:D_2},
we obtain the desired expression \eqref{eq:int_JialphaR} for the integral 
$\cal D$ in item (i).

\smallskip

\noindent
The last assertion of Lemma \ref{lem:con_balance} is an immediate consequence of 
Corollary \ref{cr:JiR} about the reactive collision operators. 
\hfill$\square$

\medskip


\begin{rem}
\label{rem:new1}
(a) The reaction rate ${\cal J}_i$ given in \eqref{eq:int_JialphaR}
can be written in an equivalent form, as a phenomenological law
for the chemical reaction \eqref{eq:reac}, as
\begin{equation}
{\cal J}_i = - \lambda_i  
             \Big( k_f  \, c^\alpha_1 c^\alpha_2 - k_b  \, c^\alpha_3 c^\alpha_4 \Big) ,
             \quad i=1,\ldots,4,
\label{eq:rr}
\end{equation}
with $k_f$ and $k_b$ being the forward and backward rate constants
given, respectively, by
\begin{equation}
k_f = \sigma^2_{12}\sqrt{\frac{8\pi k_B{T}}{\mu_{12}}}
         \exp\left(- \frac{\zeta_1}{k_B{T}}\right)
         \quad \mbox{and} \quad
k_b = \sigma^2_{34}\sqrt{\frac{8\pi k_B{T}}{\mu_{34}}}
         \exp\left(- \frac{\zeta_3}{k_B{T}}\right) ,
\label{eq:kfkb}
\end{equation}
%
\\
Equation \eqref{eq:rr} expresses the reaction rate in the form used, in general, in physical applications,
see \cite{GK10}.

\medskip

\noindent
(b) 
Our expression \eqref{eq:int_JialphaR} differs from the corresponding one obtained in paper \cite{BisiD06}, 
see (42)    
in that paper, essentially because the SRS reactive cross sections are of hard-sphere type and
the integral over $\Splus$ is explicitly evaluated in \eqref{eq:angular_int},
whereas reactive cross sections in paper \cite{BisiD06} are of Maxwell molecules type
and the integral over $\Splus$ is not explicitly evaluated.
Moreover, our exponent $1/2$, instead of $3/2$ as in paper \cite{BisiD06},
is a consequence of the fact that the exponent of the term $\big({\mu_{ij}}/{\mu_{kl}} \big)$ 
in our reactive collision operator $J_i^R$ is $2$, see \eqref{eq:JiR},
while in \cite{BisiD06} it is $3$.
\end{rem}  

\medskip


\subsection{Momentum balance equation}
\label{ssec:mbe}

The momentum balance equations for the species in the reactive mixture
can formally be derived from the scaled equations \eqref{eq:scaled_KE},
after multiplying by the molecular velocity ${\bf v}_i$ 
and then integrating over the velocity space.
Also, the conservation laws and some of the fundamental properties stated in Section \ref{sec:srs}
will be used to compute explicitly the production terms appearing in the balance equations.

However, some words are needed before presenting the next lemma.
As it is well known, the computation of the integral contributions appearing in 
the momentum balance equations is rather technical and extremely intricate,
and the final explicit expressions of these contributions are quite huge.
On the other hand, as it will become clear from the 
balance equations derived in the lemma (see \eqref{eq:momentum_balance_equation}, below), 
only the $O(1)$ terms in $\alpha$ will be retained in the equations,
that is only those terms associated to the elastic scattering will influence 
the final formulation of the balance equations.
Accordingly, we include in the next lemma only the explicit expressions of the 
$O(1)$ terms and present in Appendix \ref{App:mbe}
the explicit expression of the $O(\alpha^2)$ terms. 
%

\medskip

\begin{lem}
\label{lem:momentum_balance}
The momentum balance equation for each of the species in the reactive mixture is given by
\begin{equation}
\alpha^2\frac{\partial }{\partial t}(c_i^\alpha{\bf{u}}_i^{\alpha}) +
              \frac{k_BT}{m_i}   {\frac{\partial c_i^\alpha}{\partial {\bf{x}}}}  + 
              \alpha^2   {\frac{\partial }{\partial {\bf{x}}}}  
              \left ( c_i^\alpha{\bf{u}}_i^\alpha \otimes {\bf{u}}_i^\alpha \right )
              = {\cal O}_i - {\cal P}_i + {\cal Q}_i ,
\label{eq:momentum_balance_equation}
\end{equation}
where ${\cal O}_i$ is an $O(1)$ production term 
associated to the elastic scattering, and is
given by
\begin{align}
{\cal O}_i & = \frac{1}{\alpha} \; \int_{\mathbb{R}^3} \! {\bf{v}}_i J_i^{b\alpha E}d{\bf{v}}_i  
                         = \frac{32}{3} \sum_{\substack{s=1\\ s\neq i}}^{4}
                            \sigma^2_{is} \; \frac{m_s}{m_i+m_s} \left( \frac{2\pi k_BT}{\mu_{is}} \right)^{\!\!\!\!\frac{1}{2}}
                            c_i^\alpha c_s^\alpha \big( {\bf{u}}_s^\alpha - {\bf{u}}_i^\alpha \big),
                            \label{eq:bispecies_momentum_balance_equation}
\end{align}
and ${\cal P}_i$, ${\cal Q}_i$ are $O(\alpha^2)$ production terms
associated to the chemical process, and are respectively given by
\begin{equation}
{\cal P}_i = \alpha \int_{\mathbb{R}^3} \! {\bf{v}}_i J_i^{b\alpha * E}d{\bf{v}}_i , \qquad
{\cal Q}_i = \alpha \int_{\mathbb{R}^3}{\bf{v}}_iJ_i^{\alpha R}d{\bf{v}}_i ,
\label{eq:momcorreactive}
\end{equation}
whose explicit expressions are given in Appendix \ref{App:mbe}, 
see \eqref{eq:bispecies_correction_momentum_balance_equation} 
and \eqref{eq:last}.
\end{lem}

\medskip


\noindent
{\it Proof.} 
First, we multiply both sides of the scaled SRS equations given in \eqref{eq:scaled_KE} 
by ${\bf{v}}_i$ and integrate with respect ${\bf{v}}_i\in {\mathbb{R}^3}$,
to obtain
\begin{multline}
\alpha\frac{\partial }{\partial t} 
                    \left( \int_{\mathbb{R}^3}{\bf{v}}_i f_i^\alpha d{\bf{v}}_i \right) 
                    +
{\frac{\partial }{\partial {\bf{x}}}}
                    \left( \int_{\mathbb{R}^3}{\bf{v}}_i\otimes {\bf{v}}_if_i^\alpha d{\bf{v}}_i\right )
                    =
\frac{1}{\alpha}\int_{\mathbb{R}^3}{\bf{v}}_i J_i^{m\alpha E}d{\bf{v}}_i
\\
+
\underbrace{\frac{1}{\alpha}\int_{\mathbb{R}^3}{\bf{v}}_i J_i^{b\alpha E}d{\bf{v}}_i}_{{O}_i}
-
\underbrace{\alpha \int_{\mathbb{R}^3}{\bf{v}}_iJ_i^{b*\alpha E}d{\bf{v}}_i}_{{{\cal P}_i}}+
\underbrace{\alpha \int_{\mathbb{R}^3}{\bf{v}}_iJ_i^{\alpha R}d{\bf{v}}_i}_{{{\cal Q}_i}}.
\label{eq:momentum_balance1}
\end{multline}


\noindent
Let us  concentrate first on the left-hand-side  terms in 
\eqref{eq:momentum_balance1}.

\noindent
(i) 
From the second expression in \eqref{eq:local_max_moments}, 
it immediately follows that
\begin{equation}
\alpha\frac{\partial }{\partial t} 
                    \left( \int_{\mathbb{R}^3}{\bf{v}}_i f_i^\alpha d{\bf{v}}_i \right) 
                    =\alpha^2\frac{\partial }{\partial t} \Big(c_i^\alpha{\bf{u}}_i^{\alpha} \Big).
\label{eq:L}
\end{equation}

\medskip

\noindent
(ii) 
For what concerns the second term on the left-hand side of \eqref{eq:momentum_balance1},
we transform from ${\bf{v}}_i$ to the peculiar velocity
$\boldsymbol{\xi}_i={\bf{v}}_i-\alpha{\bf{u}}_i^\alpha$ and then use the fact that the
Jacobian of the transformation is equal to $1$ to obtain
\begin{align}
{\frac{\partial }{\partial {\bf{x}}}}
     \left( \int_{\mathbb{R}^3}{\bf{v}}_i\otimes {\bf{v}}_if_i^\alpha d{\bf{v}}_i\right )
     &=
     {\frac{\partial }{\partial {\bf{x}}}}
     \left ( 
     \int_{\mathbb{R}^3}  \!
     f_i^\alpha \; \boldsymbol{\xi}_i\otimes\boldsymbol{\xi}_i \; d\boldsymbol{\xi}_i \right )
     +
     2\alpha  {\frac{\partial }{\partial {\bf{x}}}}  
     \left ( {\bf{u}}_i^{\alpha}\otimes\int_{\mathbb{R}^3}
     f_i^\alpha \boldsymbol{\xi}_i  d\boldsymbol{\xi}_i \right )
     \nonumber
     \\
     &+
     \alpha^2  {\frac{\partial }{\partial {\bf{x}}}}  
     \left ( {\bf{u}}_i^{\alpha}
     \otimes{\bf{u}}_i^{\alpha} \int_{\mathbb{R}^3}  
     f_i^\alpha d\boldsymbol{\xi}_i \right ).
\label{eq:M}
\end{align}
Inserting \eqref{eq:local_max1}, 
expressed in terms of the peculiar velocity $\boldsymbol{\xi}_i$,
into \eqref{eq:M}, 
we easily see that the second term on the right hand side vanishes. 
Consequently, \eqref{eq:M} reduces to
\begin{equation}
{\frac{\partial }{\partial {\bf{x}}}}
     \left( \int_{\mathbb{R}^3}{\bf{v}}_i\otimes {\bf{v}}_if_i^\alpha d{\bf{v}}_i\right )
     =
     \frac{k_BT}{m_i}  {\frac{\partial c_i^\alpha}{\partial {\bf{x}}}} 
     +
     \alpha^2  {\frac{\partial }{\partial {\bf{x}}}}  
     \big( c_i^\alpha{\bf{u}}_i^\alpha \otimes {\bf{u}}_i^\alpha \big). 
\label{eq:M_final}
\end{equation}


\noindent
Now let us deal with the terms on the right hand side of 
\eqref{eq:momentum_balance1}.

\medskip

\noindent
(iii) 
The first term vanishes by virtue of the second assertion 
in Corollary \ref{cor:monospecies_MMKE_con} about the mono-species elastic operator,
see \eqref{eq:Mconservation_MMKE}. 

\medskip

\noindent
(iv) 
To derive an explicit expression for the production term ${\cal O}_i$, we
use the considered approximation of $f_i^\alpha$
in the form \eqref{eq:local_maxapp} or \eqref{eq:local_maxapp1}.
Taking into account Lemma \ref{lem:weakform_BS}
with $\varphi({\bf{{v}}}_i')={\bf{{v}}}_i'$ and $\varphi({\bf{{v}}}_i)={\bf{{v}}}_i$,
we obtain
\begin{equation*}
{{\cal O}}_i = \frac{1}{\alpha}\sum_{\substack{s=1\\ s\neq i}}^{4} 
         \sigma^2_{is} \int_{\mathbb{R}^3} \int_{\mathbb{R}^3} \int_{\mathbb{S}_+^2} 
         \big( {\bf{v}}_i'-{\bf{v}}_i \big) f_i^\alpha f_s^\alpha 
         \left \langle \epsilon ,{\bf{v}}_i-{\bf{v}}_s \right \rangle 
         d\epsilon\, d{\bf{v}}_s\,d{\bf{v}}_i.
\end{equation*}
Using the first expression in \eqref{eq:EPCV} for  ${\bf{v}}_i'$, 
evaluating the integral over the sphere $\mathbb{S}_+^2$
by transforming to spherical coordinates as described in Remark \ref{rm:sphe},
and transforming the remaining sixfold integral in  ${\bf{v}}_i$ and ${\bf{v}}_s$
 to the relative velocity ${\bf{V}}^* = {\bf{v}}_i - {\bf{v}}_s$ and 
centre of mass velocity 
     ${\bf{X}}^* = (m_i{\bf{v}}_i+m_s{\bf{v}}_s) / {M_{is}}$,
     {with $M_{is} \!=\! m_i\!+\!m_s$, as defined before,}
we obtain
\begin{align}
                 {{\cal O}}_i = \frac{4\pi}{3\alpha}& \sum_{\substack{s=1\\ s\neq i}}^{4} 
                 \sigma^2_{is}                
                 \frac{m_s}{{M_{is}}}
                 c_i^\alpha c_s^\alpha \frac{(m_im_s)^{\frac{3}{2}}}{(2\pi k_BT)^3}
                 \hspace*{1cm}
                 \label{eq:intD}
                 \\
                 \times & \Bigg\{ \!\!
                 \left( \int_{\mathbb{R}^3} \!\!
                {{V}^*} \!\! \exp\left( \!\! - \frac{\mu_{is}V^{*2}}{2k_BT} \right) \!
                {\bf{V}}^* d{\bf{V}}^* \right)             
                \left( \int_{\mathbb{R}^3}   
                \exp\left( \! - \frac{M_{is}X^{*2}}{2k_BT} \right)
                d{\bf{X}}^* \right)
                \nonumber 
                \\                 
                &
                + (a_s+a_i) 
                 \left( \int_{\mathbb{R}^3} \!\! V^*  
                \exp\left( \! - \frac{\mu_{is}V^{*2}}{2k_BT} \right) \!
                {\bf{V}}^{*}d{\bf{V}}^* \! \right)  \!\!           
                 \left( \int_{\mathbb{R}^3} \!\!
                \exp\left( \! - \frac{M_{is}X^{*2}}{2k_BT} \right) \!
                {\bf{X}}^* d{\bf{X}}^* \! \right)                                               
                \nonumber
                \\
                & 
                + \frac{a_sm_i - a_im_s}{M_{is}}  
                 \left( \int_{\mathbb{R}^3}   
                \exp\left( \! - \frac{\mu_{is}V^{*2}}{2k_BT} \right) \!
                {{V}}^{*3} d{\bf{V}}^* \right) \!             
                 \left( \int_{\mathbb{R}^3} 
                \exp\left( \! - \frac{M_{is}X^{*2}}{2k_BT} \! \right) \!
                d{\bf{X}}^* \right) \!\!\!             
                \Bigg \} ,
                \nonumber  
\end{align}
where we have introduced the notation 
$V^* \!=\! \left \| {\bf{V}}^* \right \|$, $X^*  \!=\!  \left \| {\bf{X}}^* \right \|$
and also
$a_i \!=\! \frac{\alpha m_i{{\bf{u}}_i^\alpha}}{k_BT}$,
$a_s \!=\! \frac{\alpha m_i{{\bf{u}}_s^\alpha}}{k_BT}$
as defined in \eqref{eq:local_maxapp1}.

\bigskip

\noindent
Considering the first integral in \eqref{eq:intD},
writing ${\bf{V}}^*=V^*{\bf\hat{v}}$ with ${\bf\hat{v}}=({\bf{\hat{x}}} , {\bf{\hat{y}}} , {\bf{\hat{z}}})$ a unit vector,
and transforming to spherical coordinates,
we conclude that the integral va\-ni\-shes,
since
\begin{align*}
\int_{\mathbb{R}^3} \!\!\!\!
       {{V}^*} \!\! \exp \! & \left( \!\! - \frac{\mu_{is}V^{*2}}{2k_BT} \right) \!\!  {\bf{V}}^* \!\! d{\bf{V}}^* 
       =
       \int_0^\infty \!\!\!\! {{V}^{*4}} \exp \! \left( \!\! - \frac{\mu_{is} V^{*2}}{2k_BT} \right) \! d{{V}}^*
       \Bigg( \!
       \int_0^\pi \!\! \sin^2\theta \, d\theta \!\! \int_0^{2\pi} \!\! \cos\phi d\phi \; {\bf{\hat{x}}}
       \\
       & \qquad\qquad +
       \int_0^\pi \!\!  \sin^2\theta \,d\theta \!\! \int_0^{2\pi} \!\! \sin\phi\, d\phi \; {\bf{\hat{y}}} +
       \int_0^\pi \!\! \sin\theta \cos\theta \,d\theta \!\! \int_0^{2\pi} d\phi \; {\bf{\hat{z}}}  \Bigg)
       =0 .
\end{align*}
This implies that the first two addends within the braces in \eqref{eq:intD} vanish.
Concerning now the integrals in the last addend of the same equation, 
we have that the integral in ${\bf{X}}^*$ is similar to the one in \eqref{eq:D_1_1},
whereas the integral in ${\bf{V}}^*$,
after transforming to spherical coordinates, 
results in
\begin{align*}
\int_{\mathbb{R}^3}   
\exp\left( \! - \frac{\mu_{is}V^{*2}}{2k_BT} \right) \! {{V}}^{*3} d{\bf{V}}^*
                = 4\pi\int_0^\infty V^{*5} \exp\left( \! - \frac{\mu_{is}V^{*2}}{2k_BT} \right) \! d{{V}}^*.
\end{align*}
Performing another transformation,
defined by
$ \; g \!=\! \left( \frac{\mu_{is}}{2k_BT} \right)^{\!\!\frac{1}{2}}V^*$,
and integrating,
we obtain
\begin{equation}
\int_{\mathbb{R}^3}   
\exp\left( \! - \frac{\mu_{is}V^{*2}}{2k_BT} \right) \! {{V}}^{*3} d{\bf{V}}^*
                = 
                4\pi\left( \frac{2k_BT}{\mu_{is}} \right)^{\!3} .
\label{eq:O_3_2}
\end{equation}
Substituting the above results 
into \eqref{eq:intD} and performing a little algebra
we derive the final expression \eqref{eq:bispecies_momentum_balance_equation}.
This ends the proof of Lemma \ref{lem:momentum_balance}.
The computation of the $O(\alpha^2)$ terms ${\cal P}_i$ and ${\cal Q}_i$ 
is omitted here, see Appendix \ref{App:mbe}.
\hfill$\square$

\medskip


\begin{rem}
\label{rem:new2}

\noindent
(a) 
The diffusion coefficients in our limiting equations
can be computed explicitly from expressions
\eqref{eq:bispecies_momentum_balance_equation}.
This will be done in the next subsection, 
see expression \eqref{eq:diffcoeff}.

\medskip

\noindent
(b) In paper \cite{KPBS06}, for another kinetic model,
the authors consider an input function with the same contribution with respect to the diffusion velocity
and obtain a production term similar to our term \eqref{eq:bispecies_momentum_balance_equation}, 
see (34) in that paper.
The coefficient $32/3$ in our case, instead of $8/3$ in paper \cite{KPBS06},
results from the definition of the elastic cross sections,
namely we use $\sigma_{is}^2 \!=\! (d_i + d_s)^2/4$ and
paper \cite{KPBS06} uses $\sigma_{\beta\alpha} \!=\! (d_\alpha + d_\beta)^2/16$.
\hfill$\square$
\end{rem}  

\medskip


\subsection{Macroscopic equations and formal asymptotics}
\label{ssec:mefa}

In this subsection we state our main result in this paper. 
In particular, using Lemmas \ref{lem:con_balance} and \ref{lem:momentum_balance},
we formally derive the reaction diffusion system of MS type as the hydrodynamic asymptotic limit 
of the scaled SRS kinetic system \eqref{eq:scaled_KE}.
The connection between the two systems is based
on the fact that the scaled Maxwellians \eqref{eq:local_max1}
solve the kinetic equations if the macroscopic parameters $c_i^\alpha$ and ${\bf{u}}_i^\alpha$
characterizing such Maxwellians solve the approximate equations \eqref{eq:lcbe} and \eqref{eq:momentum_balance_equation}.
The conclusion is obtained 
in the limit as $\alpha \to 0$ by assuming that the approximate functions
$c_i^\alpha$, ${\bf{u}}_i^\alpha$ converge pointwise to $c_i$, ${\bf{u}}_i$,
for $t\geq0$ and ${\bf x}\in\Omega$.

\medskip


\begin{thm}
\label{th:main}
(i) The Maxwellians defined in \eqref{eq:local_max1} are solutions of the initial boundary value problem 
for the scaled SRS kinetic equations \eqref{eq:scaled_KE} 
with initial conditions \eqref{eq:local_max}
if the parameters $c_i^\alpha$ and ${\bf{u}}_i^\alpha$ solve the approximate system 
\begin{equation}
\left\{
\begin{aligned}
      &    \frac{\partial c_i^\alpha}{\partial t}+ \frac{\partial }{\partial {\bf{x}}}( c_i^\alpha {\bf{u}}_i^\alpha)
            = {\cal J}_i ,   \\
      &    \alpha^2\frac{\partial }{\partial t}(c_i^\alpha{\bf{u}}_i^{\alpha}) + 
            \frac{k_BT}{m_i} \frac{\partial c_i^\alpha}{\partial {\bf{x}}} +\alpha^2
            \frac{\partial }{\partial {\bf{x}}} 
            \left ( c_i^\alpha{\bf{u}}_i^\alpha \otimes {\bf{u}}_i^\alpha \right )
            =  {\cal O}_i - {\cal P}_i + {\cal Q}_i,
\end{aligned}
\right.
\label{eq:con_momentum_app}
\end{equation} 
where the reaction rate ${\cal J}_i$
and the production terms ${\cal O}_i$, ${\cal P}_i$ and ${\cal Q}_i$,
defined respectively by \eqref{eq:int_JialphaR}, \eqref{eq:bispecies_momentum_balance_equation} and 
\eqref{eq:momcorreactive},
are computed using the approximations \eqref{eq:local_maxapp} to the Maxwellians \eqref{eq:local_max1}.

\bigskip

\noindent
(ii) Moreover, in the limit as $\alpha\rightarrow 0$, the system \eqref{eq:con_momentum_app} 
reduces to
\begin{equation}
\left\{
\begin{aligned}
&   \frac{\partial c_i}{\partial t} + \frac{\partial {\bf{J}}_i }{\partial {\bf{x}}}
     = \sigma^2_{ij} \left( \frac{2\pi\mu_{ij}}{ k_BT} \right)^{\!\!\!\frac{1}{2}}
        \left [ \left( \frac{\mu_{ij}}{\mu_{kl}} \right)^{\!\!\!\frac{1}{2}} \!\! c_k\,c_l
        \; \exp\! \left( \frac{Q_R}{k_B T}\right) - c_i \, c_j \right ]
        \\
&   \qquad\qquad\qquad\qquad\qquad
     \times\left [ \frac{2k_BT}{\mu_{ij}} \Gamma (2,z_i^*) -
     \left ( \frac{\Xi_{ij}}{c_0} \right )^{\!\!\!2}\Gamma (1,z_i^*) \right] ,
     \\[2mm]
&   \frac{\partial  c_i}{\partial\bf x} 
     = \frac{32}{3}\sum_{\substack{s=1\\ s\neq i}}^{4}\sigma^2_{is}
         \left( \frac{2\pi\mu_{ij}}{ k_BT} \right)^{\!\!\!\frac{1}{2}} 
         \Big( c_i{\bf{J}}_s - c_s{\bf{J}}_i \Big) ,
\label{eq:con_and_momentum_balance_equation_1}
\end{aligned}
\right.
\end{equation}
with $z_i^*$ given by \eqref{eq:zzii},
whose unknowns are the concentrations $c_i$ and the diffusive fluxes ${\bf{J}}_i$.
\end{thm}

\medskip

\noindent
{\it Proof.} The statement in item (i) follows from Lemmas \ref{lem:con_balance} and \ref{lem:momentum_balance}. 

\medskip

\noindent
Concerning item (ii), from definitions \eqref{eq:ciNigammaiji},
we can write
$c_i^\alpha{\bf{u}}_i^\alpha = {\bf{J}}_i^\alpha + c_i^\alpha {\bf{u}}^\alpha$,
where ${\bf{u}}^\alpha$ represents the molar average velocity of the mixture
relative to the scaled distributions $f_i^\alpha$, see expression \eqref{subeq:b}.
Letting $\alpha\rightarrow 0$ in \eqref{eq:con_momentum_app}, 
assuming that the limits 
\begin{equation*}
c_i = \lim_{\alpha\rightarrow 0}c_i^\alpha,    \qquad 
{\bf{J}}_i = \lim_{\alpha\rightarrow 0}{\bf{J}}_i^\alpha,  \qquad  
{\bf{u}} = \lim_{\alpha\rightarrow 0}{\bf{u}}^\alpha, 
\end{equation*}
exist pointwise for any $t>0$ and ${\bf x}\in\Omega$,
and neglecting the convective term, that is $\frac{\partial}{\partial \bf x}(c_i {\bf u})=0$,
we obtain the desired system \eqref{eq:con_and_momentum_balance_equation_1}.
\hfill $\square$


\medskip


\noindent
Observe that summing the first and second equations of \eqref{eq:con_and_momentum_balance_equation_1} 
over all species, we obtain that $c$ is uniform in space and constant in time.
Therefore, the second equation 
of system \eqref{eq:con_and_momentum_balance_equation_1} 
can be rewritten as    
\begin{equation}
\frac{\partial  c_i}{\partial\bf x} = \frac{1}{c}\sum_{\substack{s=1\\ s\neq i}}^{4}\frac{ c_i{\bf{J}}_s - c_s{\bf{J}}_i }{D_{is}} \;,
\label{eq:newnew}
\end{equation}
where $ D_{is}$ are the diffusion coefficients given by     
\begin{equation}
D_{is} = \frac{3}{32} \left( \frac{k_BT}{2\pi\mu_{is}} \right)^{\!\!\frac{1}{2}}\frac{1}{c \, \sigma_{is}^2} .
\label{eq:diffcoeff}
\end{equation}
Consequently, for $i=1,2,3,4$, putting together the constraint law \eqref{eq:Jsum}
the boundary conditions  \eqref{eq:BDcondition} for the diffusive fluxes ${\bf J}_i$,
and \eqref{eq:con_and_momentum_balance_equation_1}
with the gradient term $ \frac{\partial c_i}{\partial {\bf{x}}}$ expressed by \eqref{eq:newnew},
we obtain 
\begin{equation}
\begin{aligned}
\left\{\begin{array}{r@{\mskip\thickmuskip}l}
     & \displaystyle \frac{\partial c_i}{\partial t} +
     \frac{\partial {\bf {J}}_i}{\partial {\bf{x}}} 
     = {\cal J}_i ,
     \quad {\bf{x}}\,\in\,\Omega, \quad t>0,
     \\[1em]
     & \displaystyle \frac{\partial c_i}{\partial {\bf{x}}}
     = - \frac1c \displaystyle\sum_{\substack{s=1\\ s\neq i}}^{4}
     \frac{c_s{\bf {J}}_i-c_i{\bf {J}}_s}{D_{is}} \;,
     \quad {\bf{x}}\,\in\,\Omega, \quad t>0,\\
     & \displaystyle \sum_{i=1}^{4}{\bf {J}}_i = 0 ,
     \quad {\bf{x}}\,\in\,\Omega,\quad t>0 ,
     \\[1.5em]
     & \displaystyle \nu \cdot {\bf{J}}_i = 0 ,
     \qquad {\bf{x}}\,\in\,\partial\Omega,\quad t>0.
\end{array} \right.
\end{aligned}
\label{eq:MS_reaction_diffusion_equation}
\end{equation}   
System \eqref{eq:MS_reaction_diffusion_equation} constitutes a boundary value problem 
which we refer to as the reaction diffusion system of Maxwell-Stefan type.            
                   
\medskip

                
\section{Conclusion}  
\label{sec:conc}      
                    
In this paper, we formally derive a reaction diffusion system of Mawell-Stefan type
as the hydrodynamic limit of the simple reacting spheres kinetic model for a quaternary mixture of monatomic ideal gases
undergoing a reversible chemical reaction of bimolecular type.
By considering a scaling in which elastic collisions play a dominant role in the evolution of the species 
while chemical reactions are slow, and using a first order correction to the Maxwellian distribution in the species rest frame, 
the diffusion coefficients and the chemical production rates appearing in the species equations for the 
concentration and momentum have been explicitly computed from the collisional dynamics of the kinetic model.

\medskip

An important aspect in our work is that we have used the same correction to the Maxwellian distribution 
for the computation of both elastic and reactive production terms,
leading to a more consistent macroscopic picture.
Moreover, our correction to the Maxwellian distribution coincides with the one used by Stefan 
in his celebrated work from 1872
to derive the diffusion coefficients in a mixture of gases.

\medskip

In the quoted literature about the hydrodynamic limit of a kinetic model for reactive mixtures, 
the derivation of the MS equations had not yet been considered.
Our work provides the first result in this direction and, in our opinion, complements the work developed 
in  \cite{BoudinGS12, BoudinGS15, BGP17, HS17} in the context of non-reactive mixtures.

\medskip

Still in connection with the work developed in \cite{BoudinGS12, BoudinGS15, BGP17, HS17}
for non-reactive mixtures, we would like to emphasize here that
when we turn off the chemical reaction in our model (i.e if the mixture is made up of four non-reactive species), 
our limiting system \eqref{eq:MS_reaction_diffusion_equation} reduces to the Maxwell-Stefan system 
for hard sphere molecules, 
which is similar to the one obtained in \cite{BoudinGS15} for Maxwellian molecules. 

\medskip

The fact that in our analysis the mixture can react chemically
allows to consider many interesting problems concerning the derivation of the MS equations from the kinetic model.
Among the most interesting problems, we quote the following.

\medskip

The first problem is the introduction of the chemical potentials as the main agent 
in the definition of the driving forces and the study of the passage to the hydrodynamic limit
by removing the isothermal assumptions.
This will certainly leads to a very rich setting from the physical but also from the mathematical point of view.

\medskip

The second problem concerns the possibility of studying different time scales associated to the chemical reaction,
in particular different chemical regimes,
and obtaining the influence of the chemical reaction in the limiting MS equations.
In fact, in the case studied here, the contribution coming from the reactive collision terms
do not give any contribution to the limit equations for the momentum of the species,
see Lemma \ref{lem:momentum_balance}.

Adionally, when studying different time scales associated to the chemical reaction,
especially those chemical regimes in which reactive collisions are treated in equal pair with elastic ones, 
our SRS model results to be appropriate, because the inclusion of the correction term 
in the elastic operator prevents double counting of certain collisions,
see the explanations about in the elastic operator defined in \eqref{eq:JiE}.
In this sense, the correction term leads to non anomalous results.

\medskip

The problems just described above will be addressed in future works. 

\bigskip


\noindent
{\bf Acknowledgments.}
B.A. and A.J.S. thank Centro de Matem\'atica da Universidade do Minho, Portugal,
and the FCT/Portugal Project \! UID/MAT/00013/2013. \!
B.A. thanks the  FCT/Portugal  for the support through the PhD grant PD/BD/128188/2016.
P.G. thanks FCT/Portugal  for the support through the project 
UID/MAT/04459/2013 and the 
French Ministry of Education through the grant ANR (EDNHS).
The authors thank the Program Pessoa of Cooperation between 
Portugal and France with reference 406/4/4/2017/S.


\bigskip


\section{Appendices}
\label{sec:App}

In this section, we give some details about the computation of the integrals 
appearing in Section \ref{sec:rdl} and not evaluated there.

\subsection{Appendix A}
\label{App:cbe}

In this appendix we prove that 
${\cal E}$ and ${\cal F}$,
appearing in \eqref{eq:Jialpha_approx} of Subsection \ref{ssec:cbe} are null.  
We start with 
${\cal E}$ and we write it as the sum of the next two terms
\begin{equation*}
\begin{aligned}
{\cal E}_1={\sigma^2 _{ij} \left( \frac{\mu_{ij}}{\mu_{kl}} \right)^{\!\!2} a_k
                          \int _{{\mathbb{R}}^3}\int _{{\mathbb{R}}^3}\int _{{\mathbb{S}^2_+}}{\bf{v}}_k^{\circ}
                          M_k^\alpha M_l^\alpha \Theta \left ( \left \langle \epsilon ,{c_0\bf{v}}_i-{c_0\bf{v}}_j \right \rangle-\Xi _{ij} \right )
                          \left \langle \epsilon ,{\bf{v}}_i-{\bf{v}}_j \right \rangle d\epsilon \,d{\bf{v}}_j \,d{\bf{v}}_i} ,\\
{\cal E}_2= {\sigma^2 _{ij} \left( \frac{\mu_{ij}}{\mu_{kl}} \right)^{\!\!2} a_l\int _{{\mathbb{R}}^3}
                          \int _{{\mathbb{R}}^3}\int _{{\mathbb{S}^2_+}}{\bf{v}}_l^{\circ}M_k^\alpha M_l^\alpha 
                          \Theta \left ( \left \langle \epsilon ,{c_0\bf{v}}_i-{c_0\bf{v}}_j \right \rangle-\Xi _{ij} \right )
                          \left \langle \epsilon ,{\bf{v}}_i-{\bf{v}}_j \right \rangle d\epsilon \,d{\bf{v}}_j \,d{\bf{v}}_i}.
\end{aligned}
\end{equation*}
From \eqref{eq:max0} we rewrite the integral appearing in ${\cal E}_1$ as
\begin{multline*}
c_k^\alpha\,c_l^\alpha\frac{(m_km_l)^{\frac{3}{2}}}{(2\pi k_BT)^3}
\int _{{\mathbb{R}}^3}\int _{{\mathbb{R}}^3}\int _{{\mathbb{S}^2_+}}{\bf{v}}_k^{\circ}
\exp \left[ - \frac{m_k({\bf{v}}_k^{\circ})^2+m_l({\bf{v}}_l^{\circ})^2}{2k_BT} \right]
\\
\times\Theta \left ( \left \langle \epsilon ,{c_0\bf{v}}_i-{c_0\bf{v}}_j \right \rangle-\Xi _{ij} \right )\left \langle \epsilon ,{\bf{v}}_i-{\bf{v}}_j \right \rangle d\epsilon \,d{\bf{v}}_j \,d{\bf{v}}_i .
\end{multline*}
Now we change variables  
${\bf{v}}_i, \,{\bf{v}}_j$ to ${\bf{v}}^{\circ}_k, \, {\bf{v}}^{\circ}_l$,
and use {\eqref{eq:activationenergy3} and Property \ref{lem:Transformation}}   
to rewrite the previous term as 
\begin{multline}
c_k^\alpha\,c_l^\alpha\frac{(m_km_l)^{\frac{3}{2}}}{(2\pi k_BT)^3}   
\left( \frac{\mu_{kl}}{\mu_{ij}} \right)^{\!\!2} \int_{\mathbb{R}^3} 
\int_{\mathbb{R}^3}\int_{\mathbb{S}_+^2}{\bf{v}}^{\circ}_k\,
\exp \left[ - \frac{m_k({\bf{v}}_k^{\circ})^2+m_l({\bf{v}}_l^{\circ})^2}{2k_BT} \right]
\\
\times\Theta \left( \left \langle \epsilon, c_0{\bf{v}}_k^{\circ}-c_0{\bf{v}}_l^{\circ} \right \rangle -\Xi_{kl} \right ) 
\left \langle \epsilon, {\bf{v}}^{\circ}_k-{\bf{v}}^{\circ}_l \right \rangle\,d\epsilon\,d{\bf{v}}^{\circ}_k\,d{\bf{v}}^{\circ}_l.
\label{eq:E_11}
\end{multline}
Performing similar computations to those of \eqref{eq:angular_int} we conclude that 
integral in \eqref{eq:E_11} is equal to  the difference of the next two integrals
\begin{equation}
\begin{aligned}
& {{\cal E}_1^1}={\int_{\mathbb{R}^3}\int_{\mathbb{R}^3}{\bf{v}}^{\circ}_k\,
\exp \left[ - \frac{m_k({\bf{v}}_k^{\circ})^2+m_l({\bf{v}}_l^{\circ})^2}{2k_BT} \right]
V_* \Theta \left ( V_*-\frac{\Xi_{kl}}{c_0} \right ) d{\bf{v}}^{\circ}_k\,d{\bf{v}}^{\circ}_l} ,
\\
& {{\cal E}_1^2}={\int_{\mathbb{R}^3}\int_{\mathbb{R}^3}{\bf{v}}^{\circ}_k 
\exp \left[ - \frac{m_k({\bf{v}}_k^{\circ})^2+m_l({\bf{v}}_l^{\circ})^2}{2k_BT} \right]
V_* \Theta \left ( V_*-\frac{\Xi_{kl}}{c_0} \right )\left ( \frac{\Xi_{kl}}{c_0V_*} \right )^2  
d{\bf{v}}^{\circ}_k d{\bf{v}}^{\circ}_l},
\label{eq:E_12}
\end{aligned}     
\end{equation}
where  $V_*=\left \| {\bf{v}}_k^{\circ}-{\bf{v}}_l^{\circ} \right \|=\left \|{\bf{v}}_l^{\circ}- {\bf{v}}_k^{\circ} \right \|$. 
Now, we look first at ${{\cal E}_1^1}$. 
We transform the six fold integral in ${\bf{v}}^{\circ}_k$ and ${\bf{v}}^{\circ}_l$ 
to the centre of mass velocity 
      ${\bf{X}}_*=(m_k{\bf{v}}_k^{\circ}+m_l{\bf{v}}_l^{\circ})/{M}$
and relative velocity $\bf{V}_*$. Using the fact that the Jacobian of the transformation is $1$, we obtain
\begin{equation*}
\begin{split}
{{\cal E}_1^1}=&\int_{\mathbb{R}^3} {\bf{X}}_*\,\exp\left(-\frac{MX_*^2}{2k_BT} \right)\,d{\bf{X}}_*
\int_{\mathbb{R}^3}\exp\left(- \frac{\mu_{kl}V_*^2}{2k_BT} \right) V_* \Theta 
\left ( V_*-\frac{\Xi_{kl}}{c_0} \right )\,d{\bf{V}}_*\\
                 -&\frac{m_l}{\textcolor{black}{M}}{\int_{\mathbb{R}^3} \exp\left(- \frac{MX_*^2}{2k_BT} \right) \,d{\bf{X}}_*}{\int_{\mathbb{R}^3}{\bf{V}}_*
                 \exp\left(-\frac{\mu_{kl}V_*^2}{2k_BT} \right) V_* \Theta \left ( V_*-\frac{\Xi_{kl}}{c_0} \right )\,d{\bf{V}}_*},
                 \end{split}
\end{equation*}
where $X_*=\left \| {\bf{X}}_* \right \|$. 
Note  that the integral in $\bf{X}_*$ in the first expression above is null and so the first expression above vanishes. Moreover, the integral in $\bf{V}_*$ in the second expression above, can be written in spherical coordinates and it equals to 
                \begin{align*}
                  \int_{V_*\geq \frac{\Xi_{kl}}{c_0}}^\infty\int_0^\pi \int_0^{2\pi}{{V}}_*^4{\bf\hat{v}}
                  \exp\left(- \frac{\mu_{kl}V_*^2}{2k_BT} \right)
                  \sin\theta \,d\phi\,d\theta\,d{{V}}_* ,
                   \end{align*}
where we used the fact that any vector can be written in terms of a unit vector i.e. 
${\bf{V}}_*=V_*{\bf\hat{v}}$ with ${\bf\hat{v}}={\bf{\hat x}}\sin\theta \cos\phi+{\bf{\hat y}}\sin\theta \sin\phi +{\bf{\hat z}}\cos\theta$ and that $V_*=\left \| {\bf{V}}_* \right \|$ we get, by simple computations that it is equal to zero. 
This shows that $ {\cal E}_1^1=0$.
Now, repeating exactly the same strategy as above  in order to show that ${\cal E}_1^2=0$, it is enough to compute the next two integrals

                     \begin{equation*}
                     \begin{split}
                {{\cal E}_{1A}^{2}}=&\int_{\mathbb{R}^3}{\bf{X}}_*\,
                \exp\left(- \frac{MX_*^2}{2k_BT} \right)  \,d{\bf{X}}_*\,\int_{\mathbb{R}^3}
               \exp\left(-\frac{\mu_{kl}V_*^2}{2k_BT} \right) V_* \Theta \left ( V_*-\frac{\Xi_{kl}}{c_0} \right )
               \left ( \frac{\Xi_{kl}}{c_0V_*} \right )^2\,d{\bf{V}}_* ,\\
                 {{\cal E}_{1B}^{2}}=&{\int_{\mathbb{R}^3}{\bf{V}}_*\,
                 \exp\left(-\frac{\mu_{kl}V_*^2}{2k_BT} \right) V_* \Theta 
                 \left ( V_*-\frac{\Xi_{kl}}{c_0} \right )\left ( \frac{\Xi_{kl}}{c_0V_*} \right )^2 \,d{\bf{V}}_*}
             \int_{\mathbb{R}^3}\exp\left(-\frac{MX^2}{2k_BT}\right) \,d{\bf{X}}_* .
             \end{split}
                 \end{equation*}
Note that the integral in ${\bf{X}_*}$ in the expression of ${\cal E}_{1A}^{2}$ is null, 
so that  ${\cal E}_{1A}^{2}=0$. 
The integral in  ${\bf{V}}_*$ in the expression of  $ {\cal E}_{1B}^{2}$ can be written in spherical coordinates and by writing ${\bf{V}}_*=V_*{\bf\hat{v}}$ with ${\bf\hat{v}}$ a unit vector and $V_*=\left \| {\bf{V}}_* \right \|$ we can easily show that it vanishes,
so ${\cal E}_{1B}^{2}=0$.
Thus ${\cal E}_1^{2}=0$ and ${\cal E}_1=0$ as well.
Similar computations show that ${\cal E}_2=0$.
Putting all together we conclude that $ {\cal E}=0.$

\bigskip

\noindent
Now we prove that 
${\cal F}$ given in \eqref{eq:Jialpha_approx} is equal to zero. 
As we have done above, we  write it as the sum of the next two terms
                   \begin{equation}
                   \begin{aligned}
                     {\cal F}_1&={\sigma^2 _{ij}a_i\int _{{\mathbb{R}}^3}\int _{{\mathbb{R}}^3}\int _{{\mathbb{S}^2_+}}
                     {\bf{v}}_iM_i^\alpha M_j^\alpha   \Theta \left ( \left \langle \epsilon ,{c_0\bf{v}}_i-{c_0\bf{v}}_j \right \rangle-\Xi _{ij} \right )
                     \left \langle \epsilon ,{\bf{v}}_i-{\bf{v}}_j \right \rangle d\epsilon \,d{\bf{v}}_j \,d{\bf{v}}_i} , \\
                    {\cal F}_2&={\sigma^2 _{ij}a_j\int _{{\mathbb{R}}^3}\int _{{\mathbb{R}}^3}\int _{{\mathbb{S}^2_+}}
                    {\bf{v}}_jM_i^\alpha M_j^\alpha   \Theta \left ( \left \langle \epsilon ,{c_0\bf{v}}_i-{c_0\bf{v}}_j \right 
                    \rangle-\Xi _{ij} \right )\left \langle \epsilon ,{\bf{v}}_i-{\bf{v}}_j \right \rangle d\epsilon \,d{\bf{v}}_j \,d{\bf{v}}_i}.
                   \end{aligned}
                   \label{eq:F}
                   \end{equation}
Evaluating the integral over $\mathbb S_+^2$ in ${\cal F}_1$ and ${\cal F}_2$ using  
\eqref{eq:angular_int}, 
we obtain
\begin{equation*}
\int_{\mathbb{R}^3} \int_{\mathbb{R}^3}{\bf{v}}_i 
\exp\left[ -\frac{m_i({\bf{v}}_i)^2+m_j({\bf{v}}_j)^2}{2k_BT} \right] V \Theta \left ( V-\frac{\Xi_{ij}}{c_0} \right )\left [1 -\left ( \frac{\Xi_{ij}}{c_0V} \right )^2 \right ]\,d{\bf{v}}_j\,d{\bf{v}}_i,
                   \end{equation*}
                   \begin{equation*}
\int_{\mathbb{R}^3} \int_{\mathbb{R}^3}{\bf{v}}_j 
\exp\left[ -\frac{m_i({\bf{v}}_i)^2+m_j({\bf{v}}_j)^2}{2k_BT} \right] V \Theta \left ( V-\frac{\Xi_{ij}}{c_0} \right )\left [1 -\left ( \frac{\Xi_{ij}}{c_0V} \right )^2 \right ]\,d{\bf{v}}_j\,d{\bf{v}}_i.
                   \end{equation*}
In order to show that ${\cal F}_1$ and ${\cal F}_2$  vanish it is enough to compute the above integrals which
are quite similar to ${\cal E}_1^1$, 
so that one can perform exactly the same computations as we did above to conclude that they both vanish.
This shows that    $ {\cal F}=0$.

\subsection{Appendix B}
\label{App:mbe}

In this section we compute the integrals 
${\cal P}_i$ and ${\cal Q}_i$,
that appear  in \eqref{eq:Jialpha_approx}. 
Our first goal is to show that
\begin{align}
{\cal P}_i = \frac83 \alpha^2 \sigma^2_{ij} \sqrt{\pi} \frac{m_j}{\textcolor{black}{M}}                             
                             c_i^\alpha c_j^\alpha ({\bf{u}}_j^\alpha-{\bf{u}}_i^\alpha )                                       
                            \left[ \! \left( \! \frac{2k_BT}{\mu_{ij}} \right)^{\!\!\!\!\frac{3}{2}}\Gamma(3,z^*_i) - 
                            \left( \! \frac{\Xi_{ij}}{c_0} \! \right )^{\!\!3}  
                            \Gamma \!\left(\frac{3}{2}, z^*_i\right ) \! \right] \!.                   
\label{eq:bispecies_correction_momentum_balance_equation} 
\end{align}
From the definition of $J_i^{b\alpha*E}$ given in \eqref{eq:jib*e} and using 
Lemma \ref{lem:weakform_BS} we obtain 
\begin{equation*}
                 {\cal P}_i= \sigma^2_{ij}{\alpha}\int_{\mathbb{R}^3} \int_{\mathbb{R}^3} \int_{\mathbb{S}_+^2} 
                 \left ( {\bf{v}}_i'-{\bf{v}}_i \right) f_if_j\Theta \left ( \left \langle \epsilon ,{\bf{v}}_i-{\bf{v}}_j \right \rangle -\Xi_{ij}\right )\left \langle \epsilon ,{\bf{v}}_i-{\bf{v}}_j \right \rangle d\epsilon\, d{\bf{v}}_j \,d{\bf{v}}_i.     
                 \end{equation*}
Noting that  the equality on the left hand side of \eqref{eq:EPCV} with $s$ replaced by $j$, 
can be written as {${\bf v}_i'={\bf v}_i-2\frac{m_j}{\textcolor{black}{M}}({\bf v}_j - {\bf v}_i) \cos \theta$}. 
From this, and writing the previous integral in spherical coordinates,  
we obtain
                 \begin{multline*}
                {{\cal P}_i}=  2\sigma^2_{ij}\alpha\frac{m_j}{\textcolor{black}{M}}\\
                \quad\times\int_{\mathbb{R}^3} \int_{\mathbb{R}^3} \int_0^{\frac{\pi}{2}}\int_0^{2\pi}({\bf{v}}_j-{\bf{v}}_i)   f_i^\alpha f_j^\alpha \Theta \left ( \left \langle \epsilon ,c_0{\bf{v}}_i-c_0{\bf{v}}_j \right \rangle -\Xi_{ij}\right )V\cos^2\theta \sin\theta \,d\phi\,d\theta\, d{\bf{v}}_j\, d{\bf{v}}_i.
                 \end{multline*}
                 Integrating with respect to $\phi$ and observing that
                 \begin{align}
                 \nonumber\int_0^{\arccos\left ( \frac{\Xi_{ij}}{c_0 V} \right )} \cos^2\theta\,\sin\theta\, d\theta
                 \nonumber=\frac{1}{3}\left [ 1- \left ( \frac{\Xi_{ij}}{c_0V} \right )^3 \right ],
                 \label{eq:angular_int_correction}
                 \end{align}
              we get
                 \begin{align*}
                 {{\cal P}_i}&=\sigma^2_{ij}\alpha\frac{4\pi}{3}\frac{m_j}{\textcolor{black}{M}}\int_{\mathbb{R}^3} \int_{\mathbb{R}^3} {\bf{V}}   f_i^\alpha f_j^\alpha \Theta\left ( V-\frac{\Xi_{ij}}{c_0} \right )V\left [ 1- \left ( \frac{\Xi_{ij}}{c_0V} \right )^3 \right ]\, d{\bf{v}}_j\, d{\bf{v}}_i,
                \end{align*}
                where ${\bf{V}}$ is the relative velocity  and $V=\left \| {\bf{v}}_j-{\bf{v}}_i \right \|$. Now, we can split $ {{\cal P}_i}$ into the difference of the next two terms
                \begin{equation}
                \begin{aligned}
                  {{{\cal P}}_{i_1}}&={\alpha\sigma^2_{ij}\frac{4\pi}{3}\frac{m_j}{\textcolor{black}{M}}\int_{\mathbb{R}^3} \int_{\mathbb{R}^3} {\bf{V}}   f_i^\alpha f_j^\alpha \Theta\left ( V-\frac{\Xi_{ij}}{c_0} \right )V\, d{\bf{v}}_j\, d{\bf{v}}_i} , \\              
                 {{{\cal P}}_{i_2}}&={\sigma^2_{ij}\alpha\frac{4\pi}{3}\frac{m_j}{\textcolor{black}{M}}\int_{\mathbb{R}^3}\int_{\mathbb{R}^3} \int_{\mathbb{R}^3} {\bf{V}}   f_i^\alpha f_j^\alpha \Theta\left ( V-\frac{\Xi_{ij}}{c_0} \right )V \left ( \frac{\Xi_{ij}}{c_0V} \right )^3\, d{\bf{v}}_j\, d{\bf{v}}_i}.
                 \label{eq:P}  
                \end{aligned}
                \end{equation}
                Using \eqref{eq:local_maxapp1}, neglecting quadratic terms in the coefficients $a_i$ and changing   ${\bf{v}}_i$ and ${\bf{v}}_j$ to the centre of mass velocity ${\bf{X}}$ and relative velocity ${\bf{V}}$, we obtain
                 \begin{multline*}
                 {{{\cal P}}_{i_1}}=\frac{4\pi}{3}\sigma^2_{ij}\alpha\frac{m_j}{\textcolor{black}{M}}c_i^\alpha c_j^\alpha \frac{(m_im_j)^{\frac{3}{2}}}{(2\pi k_BT)^3}\int_{\mathbb{R}^3} \int_{\mathbb{R}^3} {\bf{V}} \exp\left[ - \frac{MX^{2}+\mu_{ij}V^{2}}{2k_BT} \right] \\ \times\left [ 1+(a_j+a_i){\bf{X}}+\left ( \frac{a_jm_i-a_im_j}{\textcolor{black}{M}} \right ){\bf{V}} \right ]
         V\Theta\left ( V-\frac{\Xi_{ij}}{c_0} \right )\, d{\bf{X}}\, d{\bf{V}}, 
                 \end{multline*}
                 \begin{multline*}
                 {{{\cal P}}_{i_2}}=\frac{4\pi}{3}\sigma^2_{ij}\alpha\frac{m_j}{\textcolor{black}{M}}c_i^\alpha c_j^\alpha \frac{(m_im_j)^{\frac{3}{2}}}{(2\pi k_BT)^3}\int_{\mathbb{R}^3} \int_{\mathbb{R}^3} {\bf{V}}\exp\left[ -\frac{MX^{2}+\mu_{ij}V^{2}}{2k_BT} \right] \\ \times \left [ 1+(a_j+a_i){\bf{X}}+\left ( \frac{a_jm_i-a_im_j}{\textcolor{black}{M}} \right ){\bf{V}} \right ]
                 \Theta\left ( V-\frac{\Xi_{ij}}{c_0} \right )V\,\left ( \frac{\Xi_{ij}}{c_0V} \right )^3 d{\bf{X}}\, d{\bf{V}} .
                 \end{multline*}
                We can split the integral appearing in ${{{\cal P}}_{i_1}}$ as the sum of 
                 \begin{equation*}
                     \begin{aligned}
                     {{{\cal P}}_{i_1}^1}=& \int_{\mathbb{R}^3}\int_{\mathbb{R}^3} {\bf{V}} \exp\left[ - \frac{MX^{2}+\mu_{ij}V^{2}}{2k_BT} \right]  V\Theta\left ( V-\frac{\Xi_{ij}}{c_0} \right )\, d{\bf{X}}\, d{\bf{V}}\\
                    {{{\cal P}}_{i_1}^2}= &(a_j+a_i)\int_{\mathbb{R}^3} \int_{\mathbb{R}^3} {\bf{V}}\exp\left[ -\frac{MX^{2}+\mu_{ij}V^{2}}{2k_BT} \right] {\bf{X}}V\Theta\left ( V-\frac{\Xi_{ij}}{c_0} \right )\, d{\bf{X}}\, d{\bf{V}}\\
                   {{{\cal P}}_{i_1}^3}= & \frac{a_jm_i-a_im_j}{\textcolor{black}{M}} \int_{\mathbb{R}^3} \int_{\mathbb{R}^3} {{V}^3}\,
                    \exp\left[ - \frac{MX^{2}+\mu_{ij}V^{2}}{2k_BT} \right] \Theta\left ( V-\frac{\Xi_{ij}}{c_0} \right )\, d{\bf{X}}\, d{\bf{V}}.   
                     \end{aligned}
                     \label{eq:P_1}
                 \end{equation*}
By using the fact that  ${\bf{V}}=V{\bf\hat{v}}$ with ${\bf\hat{v}}$ a unit vector and $V=\left \| {\bf{V}}\right \|$, we note that writing ${{\cal P}}^1_{i_1}$ in spherical coordinates, it is easy to see that  the integral with respect to $V$ vanishes.
                It is also simple to check that the integral with respect to  ${\bf{X}}$ appearing in ${{\cal P}}^2_{i_1}$ vanishes. Finally, the integral with respect to ${\bf{X}}$ appearing in 
                  ${{\cal P}}^3_{i_1}$ has been computed in \eqref{eq:D_1_1} and the integral with respect to ${\bf{V}}$ can be computed as we did for \eqref{eq:D_2n} and it equals to $2\pi\left ( \frac{2k_BT}{\mu_{ij}} \right )^3\Gamma(3,z^*_i)$.    
Putting together the previous computations and doing some algebra, we conclude that
                  \begin{equation}
                     {{\cal P}_{i_1}}=\frac{8\sqrt\pi}{3}\sigma^2_{ij}\alpha^2\frac{m_j}{\textcolor{black}{M}}c_i^\alpha c_j^\alpha \frac{\mu_{ij}}{k_BT}({\bf{u}}_j^\alpha-{\bf{u}}_i^\alpha)\left(\frac{2k_BT}{\mu_{ij}} \right)^{\!\!\frac{3}{2}}\Gamma(3,z^*_i). 
                     \label{eq:P_1final}
                  \end{equation}
Now we compute ${{\cal P}_{i_2}}$.  
Repeating exactly the same computations as we did for ${\cal P}_{i_1}$, 
we can split ${{\cal P}_{i_2}}$ into the sum of three terms. 
Two of them vanish by the same reason as above and the only one which survives is the following term
                   \begin{equation*}
                   \begin{split}
                   {{\cal P}_{i_2}^3}=&\frac{4\pi}{3}\sigma^2_{ij}\alpha\frac{m_j}{\textcolor{black}{M}}c_i^\alpha c_j^\alpha \frac{(m_im_j)^{\frac{3}{2}}}{(2\pi k_BT)^3} \frac{a_jm_i-a_im_j}{\textcolor{black}{M}}  \left ( \frac{\Xi_{ij}}{c_0} \right )^{\!3} \\
                   &\times\int_{\mathbb{R}^3} \exp \left( -  \frac{MX^{2}}{2k_BT} \right) \, d{\bf{X}}\,\int_{\mathbb{R}^3}  \Theta\left ( V-\frac{\Xi_{ij}}{c_0} \right ) \,\exp\left(-\frac{\mu_{ij}V^{2}}{2k_BT} \right)\,d{\bf{V}}.
                   \label{eq:P_2}
                   \end{split}
                   \end{equation*}
                 The integral with respect to ${\bf{X}}$ has been computed in  \eqref{eq:D_1_1} and the integral with respect to ${\bf{V}}$ can be computed as we did above and it is equal to 
$ \; 2\pi\left ( \frac{2k_BT}{\mu_{ij}} \right )^{\!\!\frac{3}{2}}\left ( \frac{\Xi_{ij}}{c_0} \right )^3\Gamma \left ( \frac{3}{2}, z^*_i\right )$.
Putting together the previous computations and doing some algebra, we conclude that
                    \begin{equation}
                    {{\cal P}_{i_2}}=\frac{8\sqrt\pi}{3} \sigma^2_{ij}\alpha^2\frac{m_j}{\textcolor{black}{M}}c_i^\alpha c_j^\alpha \frac{\mu_{ij}}{k_BT}({\bf{u}}_j^\alpha-{\bf{u}}_i^\alpha)\left ( \frac{\Xi_{ij}}{c_0} \right )^3\Gamma \left ( \frac{3}{2}, z^*_i\right ) .
                    \label{eq:P_2final}
                    \end{equation}
Substituting \eqref{eq:P_1final} and \eqref{eq:P_2final} into \eqref{eq:P} gives
the desired expression \eqref{eq:bispecies_correction_momentum_balance_equation}.

\medskip

\noindent
The last term on the right hand side of \eqref{eq:momentum_balance_equation},
that is the integral ${\cal Q}_i$ given in \eqref{eq:momcorreactive},
is similar to the one splitted in $\eqref{eq:Jialpha_approx}$. 
We can write ${\cal Q}_i$ as the sum of the next three terms,
\begin{align}
{\cal R}   &= \alpha\sigma^2 _{ij}\int _{{\mathbb{R}}^3}\int _{{\mathbb{R}}^3}\int _{{\mathbb{S}^2_+}}
             \left( \frac{\mu_{ij}}{\mu_{kl}} \right)^{\!\!2} 
             {(M_k^{\alpha\circ} M_l^{\alpha\circ}- M_i^\alpha M_j^\alpha)} {\bf{v}}_i  
             \nonumber \\
             & \hspace*{4cm} \times
             \Theta \left ( \left \langle \epsilon ,{c_0\bf{v}}_i-{c_0\bf{v}}_j \right \rangle-\Xi _{ij} \right )
             \left \langle \epsilon ,{\bf{v}}_i-{\bf{v}}_j \right \rangle d\epsilon d{\bf{v}}_j d{\bf{v}}_i ,
             \nonumber \\
{\cal S}    &= \alpha\sigma^2 _{ij}\int _{{\mathbb{R}}^3}\int _{{\mathbb{R}}^3}\int _{{\mathbb{S}^2_+}} 
            \left(\frac{\mu_{ij}}{\mu_{kl}} \right)^{\!\!2} 
            M_k^{\alpha\circ} M_l^{\alpha\circ}(a_k{\bf{v}}_k^{\circ}+a_l{\bf{v}}_l^{\circ}){\bf{v}}_i 
            \label{eq:J_iR} \\
            & \hspace*{4cm} \times \Theta \left ( \left \langle \epsilon ,{c_0\bf{v}}_i-{c_0\bf{v}}_j \right \rangle-\Xi _{ij} \right )
            \left \langle \epsilon ,{\bf{v}}_i-{\bf{v}}_j \right \rangle d\epsilon \,d{\bf{v}}_j \,d{\bf{v}}_i, 
            \nonumber\\
{\cal T}   &=-\alpha \sigma^2 _{ij} \!\! \int _{{\mathbb{R}}^3}\int _{{\mathbb{R}}^3}\int _{{\mathbb{S}^2_+}}
            \!\!\!\! 
            M_i^\alpha \! M_j^\alpha \! (a_i{\bf{v}}_i \!+\!a_j{\bf{v}}_j) \! {\bf{v}}_i 
            \Theta \left ( \left \langle \epsilon ,{c_0\bf{v}}_i \!-\! {c_0\bf{v}}_j \right \rangle \!-\! \Xi _{ij} \right )
            \left \langle \epsilon ,{\bf{v}}_i \!-\! {\bf{v}}_j \right \rangle d\epsilon \,d{\bf{v}}_j \,d{\bf{v}}_i.
\nonumber                  
\end{align}
First we prove that ${\cal R}_i=0$. 
Using \eqref{eq:max0}, \eqref{eq:angular_int}, writing  ${\cal R}$ in terms of centre of mass velocity ${\bf{X}}$ and relative velocity ${\bf{V}}$, and expanding, we realize that ${\cal R}$ can be written as the difference of the two next integrals
                    \begin{align*}
                    \nonumber{\cal R}_1&= {\Delta} \!  \int _{{\mathbb{R}}^3}{\bf{X}}\, 
                    \exp\left( - \frac{M{{X}}^2}{2k_BT} \right) d{\bf{X}}\,\int _{{\mathbb{R}}^3} V\,\Theta \left ( V-\frac{\Xi_{ij}}{c_0} \right )\left [1 -\left ( \frac{\Xi_{ij}}{c_0V} \right )^2 \right ] \exp\left( -\frac{\mu_{ij}{{V}}^2}{2k_BT} \right)d{\bf{V}}, \\
                   {\cal R}_2&= \frac{m_j}{\textcolor{black}{M}} {\Delta} \! \int _{{\mathbb{R}}^3}\int _{{\mathbb{R}}^3}  {\bf{V}}V\Theta \left ( V-\frac{\Xi_{ij}}{c_0} \right )\left [1 -\left ( \frac{\Xi_{ij}}{c_0V} \right )^2 \right ]
                    \exp\left( - \frac{M{{X}}^2+\mu_{ij}{{V}}^2}{2k_BT} \right) d{\bf{X}} d{\bf{V}},
                   \label{eq:R}
                    \end{align*}
where
$$ 
\Delta=\pi\alpha\sigma^2 _{ij}\frac{(m_im_j)^{\frac{3}{2}} }{(2\pi k_BT)^3}\left [ \left( \frac{\mu_{ij}}{\mu_{kl}} \right)^{\!\!\frac{1}{2}} c_k^\alpha\,c_l^\alpha\,e^{\frac{Q_R}{k_B T}}-c_i^\alpha\,c_j^\alpha \right ].
$$
Since the integral with respect to {\bf{X}} in ${\cal R}_1$ is zero, we conclude that the contribution  to  
${\cal R}$ comes only from    ${\cal R}_2$, which can be rewritten as 
                 \begin{equation*}
                 \begin{aligned}
               \frac{m_j}{\textcolor{black}{M}} {\Delta} & {\int _{{\mathbb{R}}^3} \!\!
                \exp\left( - \frac{M{{X}}^2}{2k_BT} \right) d{\bf{X}}} {\int _{{\mathbb{R}}^3}{\bf{V}} \!\!
                \exp\left( - \frac{\mu_{ij}{{V}}^2}{2k_BT} \right) V\Theta \left ( V \!-\! \frac{\Xi_{ij}}{c_0} \right ) d{\bf{V}}} \\
                 -&\frac{ m_j}{\textcolor{black}{M}} {\Delta} {\int _{{\mathbb{R}}^3} \!\! \exp\left( -\frac{M{{X}}^2}{2k_BT} \right) d{\bf{X}}}
                 {\int _{{\mathbb{R}}^3} \!\! {\bf{V}}\left ( \frac{\Xi_{ij}}{c_0V} \right )^{\!\!2} \!\! \exp\left( - \frac{\mu_{ij}{{V}}^2}{2k_BT} \right) V\Theta \left ( V \!-\!\frac{\Xi_{ij}}{c_0} \right )d{\bf{V}}}.
                 \label{eq:R_2}
                 \end{aligned}     
                 \end{equation*}
Transforming both integrals in ${\bf{V}}$ above to spherical coordinates 
and using the fact that ${\bf{V}}=V{\bf\hat{v}}$ with ${\bf\hat{v}}$ a unit vector and $V=\left \| {\bf{V}} \right \|$,
we get that both integrals vanish. This shows that ${\cal R}=0$ as desired.
               Now we analyze 
                ${\cal S}_i$, which can be written as the sum of the next two terms
\begin{align}
{\cal S}_1&= \alpha {\sigma_{ij}^2 \! \left( \! \frac{\mu_{ij}}{\mu_{kl}} \! \right)^{\!\!\!2}Ê\! a_k \! 
                             \int _{{\mathbb{R}}^3} \! \int _{{\mathbb{R}}^3} \! \int _{{\mathbb{S}^2_+}} \!\!
                             {\bf{v}}_i{\bf{v}}_k^{\circ}
                             M_k^\alpha M_l^\alpha \Theta \left ( \left \langle \epsilon ,{c_0\bf{v}}_i \!-\! {c_0\bf{v}}_j \right \rangle \!-\! \Xi _{ij} \right )
                             \left \langle \epsilon ,{\bf{v}}_i \!-\! {\bf{v}}_j \right \rangle d\epsilon d{\bf{v}}_j d{\bf{v}}_i}
                             \nonumber \\
                             \label{eq:S} \\
{\cal S}_2&=\alpha {\sigma_{ij}^2 \! \left( \! \frac{\mu_{ij}}{\mu_{kl}} \! \right)^{\!\!\!2} \! a_l \!
                            \int _{{\mathbb{R}}^3} \! \int _{{\mathbb{R}}^3} \! \int _{{\mathbb{S}^2_+}} \!\!
                            {\bf{v}}_i{\bf{v}}_l^{\circ}
                            M_k^\alpha M_l^\alpha \Theta \left ( \left \langle \epsilon ,{c_0\bf{v}}_i \!-\! {c_0\bf{v}}_j \right \rangle \!-\! \Xi _{ij} \right)
                            \left \langle \epsilon ,{\bf{v}}_i \!-\! {\bf{v}}_j \right \rangle d\epsilon d{\bf{v}}_j d{\bf{v}}_i}.
                            \nonumber
\end{align}
From $\eqref{eq:PCVFR}_1$ we can split  the integral appearing in ${\cal S}_1$ as the sum of the next five terms
\begin{align*}
{\cal S}^1&={\frac{m_i}{M}\int_{\mathbb{R}^3}\int_{\mathbb{R}^3}\int_{\mathbb{S}_+^2}{\bf{v}}_i{\bf{v}}_i M_k^{\alpha\circ}M_l^{\alpha\circ} \Theta \left ( \left \langle \epsilon,c_0{\bf{v}}_i-c_0{\bf{v}}_j \right \rangle -\Xi_{ij}\right )\left \langle \epsilon,{\bf{v}}_i-{\bf{v}}_j \right \rangle\, d\epsilon\,d{\bf{v}}_j\,d{\bf{v}}_i}\\
{\cal S}^2&={\frac{m_j}{M}\int_{\mathbb{R}^3}\int_{\mathbb{R}^3}\int_{\mathbb{S}_+^2}{\bf{v}}_i{\bf{v}}_j M_k^{\alpha\circ}M_l^{\alpha\circ} \Theta \left ( \left \langle \epsilon,c_0{\bf{v}}_i-c_0{\bf{v}}_j \right \rangle -\Xi_{ij}\right )\left \langle \epsilon,{\bf{v}}_i-{\bf{v}}_j \right \rangle\, d\epsilon\,d{\bf{v}}_j\,d{\bf{v}}_i}\\
{\cal S}^3  &= {\frac{m_l}{M} \! \sqrt{\frac{\mu_{ij}}{\mu_{kl}}} \!
\int_{\mathbb{R}^3} \! \int_{\mathbb{R}^3} \! \int_{\mathbb{S}_+^2} \!\! \!\!
{\bf{v}}_i({\bf{v}}_i \!-\! {\bf{v}}_j)  \!
M_k^{\alpha\circ} \! M_l^{\alpha\circ} \! \Theta \left ( \left \langle \epsilon,c_0{\bf{v}}_i \!-\! c_0{\bf{v}}_j \right \rangle \!-\! \Xi_{ij}\right )\left \langle \epsilon,{\bf{v}}_i \!-\! {\bf{v}}_j \right \rangle  d\epsilon d{\bf{v}}_j d{\bf{v}}_i}
\\
{\cal S}^4 & = - \frac{m_l}{M}\sqrt{\frac{\mu_{ij}}{\mu_{kl}}}\int_{\mathbb{R}^3}\int_{\mathbb{R}^3}\int_{\mathbb{S}_+^2}{\bf{v}}_i M_k^{\alpha\circ}M_l^{\alpha\circ}  \\
& \hspace*{3cm} \times \Theta \left ( \left \langle \epsilon,c_0{\bf{v}}_i-c_0{\bf{v}}_j \right \rangle -\Xi_{ij}\right )\,\left \langle \epsilon,{\bf{v}}_i-{\bf{v}}_j \right \rangle\,\epsilon\,\left \langle \epsilon,{\bf{v}}_i-{\bf{v}}_j \right \rangle\, d\epsilon\,d{\bf{v}}_j\,d{\bf{v}}_i \\
{\cal S}^5&={\frac{m_l}{M}\sqrt{\frac{\mu_{ij}}{\mu_{kl}}}\int_{\mathbb{R}^3}\int_{\mathbb{R}^3}\int_{\mathbb{S}_+^2}{\bf{v}}_i M_k^{\alpha\circ}M_l^{\alpha\circ} \Theta \left ( \left \langle \epsilon,c_0{\bf{v}}_i-c_0{\bf{v}}_j \right \rangle -\Xi_{ij}\right )\left \langle \epsilon,{\bf{v}}_i-{\bf{v}}_j \right \rangle \epsilon \omega^- d\epsilon d{\bf{v}}_j d{\bf{v}}_i}.
\end{align*}
Note that the integral appearing in
${\cal S}^1$ can be rewritten as
\begin{multline*}
c_k c_l \frac{(m_km_l)^{\frac{3}{2}}}{(2\pi k_BT)^3}
\exp\left( \frac{Q_R}{k_BT} \right ) 
                \int_{\mathbb{R}^3}\int_{\mathbb{R}^3}\int_{\mathbb{S}_+^2}{\bf{v}}_i {\bf{v}}_i 
                \exp\left(- \frac{m_i({\bf{v}}_i)^2+m_j({\bf{v}}_j)^2}{2k_BT} \right) \\
                \times\Theta \left ( \left \langle \epsilon,c_0{\bf{v}}_i-c_0{\bf{v}}_j \right \rangle -\Xi_{ij}\right )\left \langle \epsilon,{\bf{v}}_i-{\bf{v}}_j \right \rangle\, d\epsilon\,d{\bf{v}}_j\,d{\bf{v}}_i.    
                 \end{multline*}
Using \eqref{eq:angular_int} and expanding the  integral, we rewrite it as
                  \begin{equation*}
                 \begin{aligned}
               {\pi }\int_{\mathbb{R}^3}\int_{\mathbb{R}^3} & {\bf{v}}_i{\bf{v}}_i \, 
               \exp\left(- \frac{m_i({\bf{v}}_i)^2+m_j({\bf{v}}_j)^2}{2k_BT} \right) V \Theta \left ( V-\frac{\Xi_{ij}}{c_0} \right )\,d{\bf{v}}_j\,d{\bf{v}}_i\\
               &-{\pi\int_{\mathbb{R}^3}\int_{\mathbb{R}^3}{\bf{v}}_i{\bf{v}}_i 
               \exp\left(- \frac{m_i({\bf{v}}_i)^2+m_j({\bf{v}}_j)^2}{2k_BT} \right) 
               V \Theta \left ( V-\frac{\Xi_{ij}}{c_0} \right )\left ( \frac{\Xi_{ij}}{c_0V} \right )^{\!\!2} 
               d{\bf{v}}_j d{\bf{v}}_i}.
                 \end{aligned}
                 \label{eq:S^1}
                \end{equation*}
In the first integral in the last expression, we change  the integral to the centre of mass velocity ${\bf{X}}$ and relative velocity ${\bf{V}}$ and then we expand it,  
so that it becomes  equal to 
\begin{align}
                \pi \int_{\mathbb{R}^3} X^2  &
                \exp\left( - \frac{MX^2}{2k_BT} \right) d{\bf{X}}{\int_{\mathbb{R}^3}V \Theta \left ( V-\frac{\Xi_{ij}}{c_0} \right )
                \exp\left( - \frac{\mu_{ij}V^2}{2k_BT} \right) d{\bf{V}}}  
                \nonumber \\
               -&{2\pi\frac{m_j}{\textcolor{black}{M}} \int_{\mathbb{R}^3} \int_{\mathbb{R}^3}{\bf{V}}{\bf{X}}
               \exp\left( - \frac{MX^2+\mu_{ij}V^2}{2k_BT} \right) V 
               \Theta \left ( V-\frac{\Xi_{ij}}{c_0} \right ) d{\bf{X}} d{\bf{V}}} 
               \label{eq:S^1*} \\    
               +& \frac{m_j^2\pi}{\textcolor{black}{M}^2}\int_{\mathbb{R}^3}  
               \exp\left( - \frac{MX^2}{2k_BT} \right) d{\bf{X}}{\int_{\mathbb{R}^3}V^3  
               \exp\left( - \frac{\mu_{ij}V^2}{2k_BT} \right) 
               \Theta \left ( V-\frac{\Xi_{ij}}{c_0} \right ) d{\bf{V}}}.
               \nonumber 
\end{align}             
Note that the integral with respect to ${\bf{X}}$ in the second term above is equal to zero, 
so that we look at the contribution of the first and third terms above.  
These terms can easily be computed and are equal, respectively, to    
$$
3\pi\left( \frac{2\pi k_BT}{M} \right)^{\!\!\! \frac{5}{2}}\left( \frac{2k_BT}{\mu_{ij}} \right)^{\!\!2} \Gamma(2,z^*_i)
\qquad \mbox{and} \qquad
\frac{m_j^2\pi}{M^2} \left( \frac{2\pi k_BT}{M} \right)^{\!\!\frac{3}{2}}2\pi\left ( \frac{2k_BT}{\mu_{ij}} \right )^{\!\!3}\Gamma(3,z^*).
$$
Thus, doing simple computations we get that the contribution of the first term in \eqref{eq:S^1*} to ${\cal S}^1 $ is given by
\begin{align}
2\alpha^2 \! \sigma_{ij}^2 \!\! \left( \! \frac{2\pi k_BT}{\mu_{kl}} \! \right)^{\!\!\!\frac{1}{2}} \!\!\!
                   {\bf{u}}_kc_kc_l \frac{m_i m_k }{M} \!
                   \exp \! \left ( \!\! \frac{Q_R}{k_BT} \!\! \right ) \!\!\!
                   \left [ \! \frac{3}{M} \; \Gamma(2,z^*_i) \!+\!
                   \frac{2}{\mu_{ij}} \! \frac{m_j^2}{\textcolor{black}{M}^2} \Gamma(3,z^*_i) \! \right ]
                   \! .
\label{eq:S^1*_final}
\end{align}
Now, for the second term in \eqref{eq:S^1*} we can do exactly the same computations as we did for the first one, 
and we see that the contribution of this term to ${\cal S}^1$ is given by 
\begin{align}
\alpha^2 \!\! \sigma_{ij}^2 \! \mu_{ij} {\bf{u}}_k \! c_k c_l \frac{m_i m_k}{M} \!
                          \exp \! \left ( \!\! \frac{Q_R}{k_BT} \!\! \right ) \!\!\!
                          \left ( \! \frac{\Xi_{ij}}{c_0} \! \right )^{\!\!2} \!\!
                          \left( \! \frac{2\pi }{k_BT\mu_{kl}} \! \right)^{\!\!\! \frac{1}{2}} 
                          \label{eq:S^1**_final}                       
                          \\
                          \times \left [ \! \frac{3}{M} \Gamma (1,z^*_i) \!+\!
                          \frac{2}{\mu_{ij}} \frac{m_j^2}{\textcolor{black}{M}^2} \Gamma \left ( 2, z^*_i \right ) \! \right ]
                          \! .
                          \nonumber
\end{align}
Putting together \eqref{eq:S^1*_final} and \eqref{eq:S^1**_final} 
we obtain that the contribution of ${\cal S}^{1}$ to ${\cal S}_1$ is 
\begin{align}
{{\cal S}^{1}} = 2 \alpha^2 \sigma_{ij}^2 \!\! &
\left( \! \frac{2\pi k_BT}{\mu_{kl}} \! \right)^{\!\!\frac{1}{2}} \!\!\!
{\bf{u}}_k  c_kc_l \frac{m_i m_k}{M}
\exp\left ( \! \frac{Q_R}{k_BT} \! \right ) \!
\left [ \! \frac{3}{M} \Gamma(2,z^*) 
+ \frac{2}{\mu_{ij}} \frac{m_j^2}{\textcolor{black}{M}^2} \Gamma(3,z^*_i) \right ]
\nonumber  \\
& - \alpha^2 \sigma_{ij}^2 \mu_{ij} {\bf{u}}_k\,c_k\,c_l\,\frac{m_i m_k}{M}
\exp\left ( \frac{Q_R}{k_BT} \right ) \left ( \frac{\Xi_{ij}}{c_0} \right )^{\!\!2}
\left( \frac{2\pi }{k_BT\mu_{kl}} \right)^{\!\!\! \frac{1}{2}} 
\label{eq:S^1_final} \\
& \times\left [ \frac{3}{M} \; \Gamma (1,z^*_i)
+ \frac{2}{\mu_{ij}}  \frac{m_j^2}{\textcolor{black}{M}^2} \Gamma\left ( 2, z^*_i \right ) \right ].
\nonumber
\end{align}
We can repeat exactly the same approach as we did to compute the contribution of 
${\cal S}^1$ to ${\cal S}_1$ for the remaining integrals 
{to obtain} 
\begin{align}
{{\cal S}^2} = \; & 2 \alpha^2 \sigma_{ij}^2
\left( \! \frac{2\pi k_BT}{\mu_{kl}} \! \right)^{\!\!\!\frac{1}{2}}
{\bf{u}}_k c_kc_l \frac{m_j m_k}{M}
\exp \! \left ( \! \frac{Q_R}{k_BT} \! \right) \!
\left [ \! \frac{3}{M} \Gamma(2,z^*_i) - \frac{2}{\textcolor{black}{M}} 
\Gamma(3,z^*_i) \right ]
\label{eq:S^2_final}  \\
& \hspace*{2cm}   - \alpha^2 \! \sigma_{ij}^2 \! \mu_{ij} {\bf{u}}_k c_kc_l \!\!
\frac{m_j m_k}{M} \exp \!\left ( \! \frac{Q_R}{k_BT} \! \right) \!
\left ( \! \frac{\Xi_{ij}}{c_0} \! \right )^{\!\!2} \!\!
\left( \! \frac{2\pi}{\mu_{kl}k_BT} \right)^{\!\!\! \frac{1}{2}}
\nonumber  \\
& \hspace*{2cm} \times\left [ \frac{3}{M} \;\Gamma (1,z^*_i)-\frac{2}{\textcolor{black}{M}} 
\Gamma\left ( 2, z^*_i \right ) \right ],
\nonumber
\\[2mm]
{{\cal S}^3} = \; & 2\alpha^2 \! \sigma_{ij}^2 {{\bf{u}}_k} c_kc_l \!
\exp \! \left ( \! \frac{Q_R}{k_BT} \! \right ) \! \frac{m_j}{\textcolor{black}{M}} \!
\left ( \! \frac{2\pi \mu_{ij}}{k_BT} \! \right )^{\!\!\! \frac{1}{2}} \!\!
\left [ \! \frac{2k_BT}{\mu_{ij}}  \Gamma(3,z^*_i)
\!-\! \left ( \! \frac{\Xi_{ij}}{c_0} \! \right )^{\!\!2}  \Gamma\left ( 2, z^*_i \right ) \right ] ,
\label{eq:S^3_final}
\\[2mm]
{{\cal S}^4} = \; & \alpha^2 \! \sigma_{ij}^2 \! \mu_{ij} 
{\bf{u}}_k \! c_kc_l
\exp \! \left ( \!\! \frac{Q_R}{k_BT} \!\! \right ) \!\! \frac{4\sqrt\pi}{3k_BT}
\frac{m_j}{\textcolor{black}{M}} \!
\left [ \!\! \left ( \! \frac{2k_BT}{\mu_{ij}} \! \right )^{\!\!\!\frac{3}{2}} \!\!
\Gamma(3,z^*_i) \!-\! \left ( \! \frac{\Xi_{ij}}{c_0 } \! \right )^{\!\!\!3} \!\!
\Gamma \! \left ( \! \frac{3}{2},z^*_i \! \right ) \! \right ] \! ,
\label{eq:S^4_final}
\\[2mm]
{{\cal S}^5} = \; & 0.
\end{align}
From these computations, we conclude that  ${\cal S}_1$ is equal to the sum of 
\eqref{eq:S^1_final}, \eqref{eq:S^2_final}, \eqref{eq:S^3_final} and \eqref{eq:S^4_final},  
that is
\begin{align}
{\cal S}_1&=
\alpha^2 \! \sigma_{ij}^2 c_k c_l 
\exp \! \left ( \! \frac{Q_R}{k_BT} \! \right ) \!\!
\Bigg\{ \! 2
\! \left( \! \frac{2\pi k_BT}{\mu_{kl}} \! \right)^{\!\!\!\frac{1}{2}} \!\!
       {\bf{u}}_k\frac{m_i m_k }{M} \! \!  \left [ \! \frac{3}{M} 
       \Gamma(2,z^*_i) \!+\! \frac{2}{\mu_{ij}}  \frac{m_j^2}{\textcolor{black}{M}^2}
       \Gamma(3,z^*_i) \! \right ]
       \nonumber \\
& - 
\left ( \! \frac{\Xi_{ij}}{c_0} \! \right )^{\!\!2} \!
\left( \! \frac{2\pi }{k_BT\mu_{kl}} \! \right)^{\!\!\!\frac{1}{2}} \!
\mu_{ij} {\bf{u}}_k \frac{m_i m_k}{M} \!
\left [ \! \frac{3}{M} \Gamma (1,z^*_i) \!+\! 
\frac{2}{\mu_{ij}} \frac{m_j^2}{\textcolor{black}{M}^2} 
\Gamma\left ( 2, z^*_i \right ) \right ] 
\nonumber \\
&
+ 2  
\! \left( \! \frac{2\pi k_BT}{\mu_{kl}} \! \right)^{\!\!\!\frac{1}{2}} \!
{\bf{u}}_k \frac{m_j m_k}{M} \! \left [ \! \frac{3}{M} \Gamma(2,z^*_i)-\frac{2}{\textcolor{black}{M}} 
\Gamma(3,z^*_i) \! \right ]
\label{eq:S_1_final}  \\
\nonumber &
-  
\left ( \frac{\Xi_{ij}}{c_0} \right )^{\!\!2} \!\!
\left( \! \frac{2\pi }{k_BT\mu_{kl}} \! \right)^{\!\!\!\frac{1}{2}} \!
\mu_{ij} {\bf{u}}_k \frac{m_j m_k}{M} \! \left [ \! \frac{3}{M}  
\Gamma (1,z^*_i) \!-\! \frac{2}{\textcolor{black}{M}} \Gamma\left ( 2, z^*_i \right ) \right ]
\\
\nonumber &
+ 2  
{{\bf{u}}_k}
\frac{m_j}{\textcolor{black}{M}} \left( \! \frac{2\pi \mu_{ij}}{k_BT} \! \right )^{\!\!\!\frac{1}{2}} \!
\left [ \!  \frac{2k_BT}{\mu_{ij}}  \Gamma(3,z^*_i)- \left ( \frac{\Xi_{ij}}{c_0} \right )^2\,\Gamma\left ( 2, z^*_i \right ) \right ]
\\
& -  
\mu_{ij}{\bf{u}}_k\frac{4\pi^{\frac{1}{2}}}{3k_BT}
\frac{m_j}{\textcolor{black}{M}} \left [ \! \left ( \! \frac{2k_BT}{\mu_{ij}} \! \right )^{\!\!\!\frac{3}{2}} \!
\Gamma(3,z^*_i) \!-\! \left ( \! \frac{\Xi_{ij}}{c_0 } \! \right )^{\!\!3} \! 
\Gamma \! \left ( \! \frac{3}{2},z^*_i \! \right ) \! \right ] 
\Bigg\} .
\nonumber
\end{align}
Similarly,
\begin{align}
\nonumber{\cal S}_2&=
\alpha^2 \! \sigma_{ij}^2 c_k c_l \!
\exp \! \left ( \! \frac{Q_R}{k_BT} \! \right )  \!
\Bigg\{ 
2 \left( \! \frac{2\pi k_BT}{\mu_{kl}} \! \right)^{\!\!\!\frac{1}{2}}  \!\!
{\bf{u}}_l\frac{m_i m_l }{M} \! \left [ \! \frac{3}{M} 
\Gamma(2,z^*_i) \!+\! \frac{2}{\mu_{ij}} \frac{m_j^2}{\textcolor{black}{M}^2}
\Gamma(3,z^*_i) \right ] 
\\
\nonumber&
-   
\left ( \! \frac{\Xi_{ij}}{c_0} \! \right )^{\!\!2} \!\!
\left( \! \frac{2\pi }{k_BT\mu_{kl}} \! \right)^{\!\!\!\frac{1}{2}} \!\!
\mu_{ij} {\bf{u}}_l\frac{m_i m_l}{M}
\left [ \! \frac{3}{M} \Gamma (1,z^*_i) \!+\! \frac{2}{\mu_{ij}} \frac{m_j^2}{\textcolor{black}{M}^2} 
\Gamma\left ( 2, z^*_i \right ) \! \right ] 
\\
&
+ 2   
\left( \! \frac{2\pi k_BT}{\mu_{kl}} \! \right)^{\!\!\!\frac{1}{2}} \!\!
{{\bf{u}}_l}\frac{m_j m_l }{M}
\left [ \! \frac{3}{M} \Gamma(2,z^*_i) \!-\! \frac{2}{\textcolor{black}{M}} \Gamma(3,z^*_i) \! \right ]
\label{eq:S_2_final} \\
\nonumber&
-  
\left ( \! \frac{\Xi_{ij}}{c_0} \! \right )^{\!\!2} \!\!
\left( \! \frac{2\pi }{k_BT\mu_{kl}} \! \right)^{\!\!\! \frac{1}{2}} \!\!
\mu_{ij} {\bf{u}}_l \frac{m_j m_l}{M}
\left [ \! \frac{3}{M} \Gamma (1,z^*_i) \!-\! \frac{2}{\textcolor{black}{M}}
\Gamma\left ( 2, z^*_i \right ) \right ]
\\
\nonumber &
+ 2  
{{\bf{u}}_l}\frac{m_j}{\textcolor{black}{M}} \!
\left ( \! \frac{2\pi \mu_{ij}}{k_BT} \! \right )^{\!\!\!\frac{1}{2}} \!\!
\left [ \! \frac{2k_BT}{\mu_{ij}} 
\Gamma(3,z^*_i) \!-\! \left ( \! \frac{\Xi_{ij}}{c_0} \! \right )^{\!\!2} \!\!
\Gamma\left ( 2, z^*_i \right ) \! \right ]\\
& - 
\mu_{ij}{\bf{u}}_l \frac{4\sqrt \pi}{3k_BT}\
\frac{m_j}{\textcolor{black}{M}}
\left [ \left ( \frac{2k_BT}{\mu_{ij}} \! \right )^{\!\!\!\frac{3}{2}} \!\!
\Gamma(3,z^*_i) \!-\! \left ( \! \frac{\Xi_{ij}}{c_0 } \! \right )^{\!\!3} \!\! 
\Gamma \left ( \! \frac{3}{2},z^*_i \right ) \! \right ]  
\Bigg\} .
\nonumber
\end{align}
Substituting \eqref{eq:S_1_final} and \eqref{eq:S_2_final} into \eqref{eq:S}, 
we obtain

\begin{align}
\nonumber {\cal S}_i&=
\alpha^2 \! \sigma_{ij}^2 \! c_k c_l \!
\exp \! \left ( \! \frac{Q_R}{k_BT} \! \right ) \!\!
\Bigg\{ \!\!\!\!
\left( \! \frac{2\pi k_BT}{\mu_{kl}} \! \right)^{\!\!\! \frac{1}{2}} \!\! \frac{2m_i}{M} \!
\left [ \! \frac{3}{M} \Gamma(2,z^*_i) \!\!+\!\! \frac{2}{\mu_{ij}} \! \frac{m_j^2}{\textcolor{black}{M}^2}
\Gamma(3,z^*_i) \! \right ] \!
\\
\nonumber&
-   
\left ( \! \frac{\Xi_{ij}}{c_0} \! \right )^{\!\!2} \!\!
\left( \! \frac{2\pi }{k_BT\mu_{kl}} \! \right)^{\!\!\! \frac{1}{2}} \!\!
\mu_{ij}\frac{m_i}{M} \! \left [ \! \frac{3}{M}
\Gamma (1,z^*_i) \!+\!  \frac{2}{\mu_{ij}} \frac{m_j^2}{\textcolor{black}{M}^2} 
\Gamma\left ( 2, z^*_i \right ) \right ]
\\
&
+ 2   
\left( \! \frac{2\pi k_BT}{\mu_{kl}} \! \right)^{\!\!\! \frac{1}{2}} \!\!
\frac{m_j}{M} \! \left [ \! \frac{3}{M} 
\Gamma(2,z^*_i) \!-\! \frac{2}{\textcolor{black}{M}}\Gamma(3,z^*_i) \! \right ]
\label{eq:S_final}
\\
\nonumber&
-   
\left ( \! \frac{\Xi_{ij}}{c_0} \! \right )^{\!\!2} \!
\left( \! \frac{2\pi }{k_BT\mu_{kl}} \! \right)^{\!\!\! \frac{1}{2}} \!\!
\mu_{ij}\frac{m_j}{M} \! \left [ \! \frac{3}{M} 
\Gamma (1,z^*_i) \!-\! \frac{2}{\textcolor{black}{M}} 
\Gamma\left ( 2, z^*_i \right ) \right ] \!
\Bigg\}
(m_k {\bf{u}}_k+m_l {\bf{u}}_l ) 
\\[2mm]
\nonumber &
+ \alpha^2 \! \sigma_{ij}^2 \! c_k c_l \!
\exp \! \left ( \! \frac{Q_R}{k_BT} \! \right ) \!\!
\Bigg\{ \!\!\!\!
\frac{2m_j}{\textcolor{black}{M}} \! \left ( \! \frac{2\pi \mu_{ij}}{k_BT} \! \right )^{\!\!\! \frac{1}{2}} \!\!
\left [ \!  \frac{2k_BT}{\mu_{ij}} 
\Gamma(3,z^*_i) \!-\! \left ( \! \frac{\Xi_{ij}}{c_0}  \! \right )^{\!\!2} \!
\Gamma\left ( 2, z^*_i \right ) \! \right ] \!
\\
\nonumber &
-  
\mu_{ij} \frac{4\sqrt \pi}{3k_BT}  \frac{m_j}{\textcolor{black}{M}} \!
\left [ \! \left ( \! \frac{2k_BT}{\mu_{ij}} \! \right )^{\!\!\!\frac{3}{2}} \!\!
\Gamma(3,z^*_i) \!-\! \left ( \frac{\Xi_{ij}}{c_0 } \right )^{\!\!3} \! 
\Gamma \left ( \frac{3}{2},z^*_i \right ) \! \right ] \!
\Bigg\}
({{\bf{u}}_k} \!+\! {{\bf{u}}_l})  .
\nonumber 
\end{align}
Now, to treat the last term on the right hand side of \eqref{eq:J_iR}
we can do exactly the same computations as we did above, 
and we get at the end that
\begin{align}
\nonumber{\cal T}_i &= 
 \alpha^2 \sigma_{ij}^2  c_i^\alpha c_j^\alpha \Bigg\{
 2m_i {\bf{u}}_i^\alpha 
\left( \! \frac{2\pi k_BT}{\mu_{ij}} \! \right)^{\!\!\! \frac{1}{2}} \!\!\!
\left [ \! \frac{3}{M} \Gamma(2,z^*_i) +
\frac{2}{\mu_{ij}} \frac{m_j^2}{\textcolor{black}{M}^2}
\Gamma(3,z^*_i) \right ]\\
\nonumber & 
- 
m_i {\bf{u}}_i^\alpha 
\left ( \frac{\Xi_{ij}}{c_0} \! \right )^{\!\!2} \!
\left( \! \frac{2\pi\mu_{ij}}{k_BT} \! \right)^{\!\!\!\frac{1}{2}} \!
\left [ \frac{3}{M} \Gamma (1,z^*_i) + \frac{2}{\mu_{ij}}  \frac{m_j^2}{\textcolor{black}{M}^2} 
\Gamma\left ( 2, z^*_i \right ) \right ]\\
\nonumber&
+ 2 
m_j{\bf{u}}_j^\alpha 
\left( \! \frac{2\pi k_BT}{\mu_{ij}} \right)^{\!\!\!\frac{1}{2}}
\left [ \! \frac{3}{M} \Gamma(2,z^*_i) - \frac{2}{\textcolor{black}{M}}\Gamma(3,z^*_i) \right ]
\\
& - 
m_j{\bf{u}}_j^\alpha 
\left ( \frac{\Xi_{ij}}{c_0} \right )^{\!\!2} \!\!
\left( \! \frac{2\mu_{ij}\pi}{k_BT} \! \right)^{\!\!\! \frac{1}{2}} \!\!
\left [ \frac{3}{M} \Gamma (1,z^*_i)-\frac{2}{\textcolor{black}{M}} \Gamma\left ( 2, z^*_i \right ) \right ]
\Bigg\} .
\label{eq:T_final}
\end{align}
Finally, since ${\cal R}_i =0$,
we obtain 
\begin{equation*}
{\cal Q}_i = {\cal S}_i + {\cal T}_i ,
\label{eq:last}
\end{equation*}
with ${\cal S}_i$ and ${\cal T}_i $ given by
\eqref{eq:S_final} and \eqref{eq:T_final},
respectively.


\end{document}